\documentclass[12pt, draftclsnofoot, onecolumn]{IEEEtran}

\usepackage[cmex10]{amsmath}
\usepackage{amssymb}
\usepackage{graphicx}
\usepackage{array,color}
\usepackage{amsmath}
\usepackage{stfloats}
\usepackage{graphicx}
\usepackage{subfigure}
\usepackage{tabularx}
\usepackage{epsfig,epsf,color,balance,cite}
\usepackage{algorithmic}
\usepackage{algorithm}
\usepackage{bm}
\usepackage{textcomp}
\usepackage{caption}
\usepackage{multirow}
\usepackage{cite}
\usepackage{enumerate}
\usepackage{cases}
\usepackage{color}
\usepackage{url}
\begin{document}
\title{On Consideration of Content Preference and Sharing Willingness in D2D Assisted Offloading}

\author{\IEEEauthorblockN{Yijin Pan, Cunhua Pan, Huiling Zhu, Qasim Zeeshan Ahmed, Ming Chen and Jiangzhou Wang, \textit{Fellow, IEEE} \\
\thanks{This work has been accepted by the IEEE JSAC Special Issue on Human-In-The-Loop Mobile Networks.
Part of this work will be presented at the IEEE ICC 2017. Y.Pan and M.Chen are with the National Mobile Communications Research Laboratory, Southeast University, Nanjing 211111, China. Email: \{panyijin,chenming\}@seu.edu.cn. C.Pan was with the National Mobile Communications Research Laboratory, Southeast University, Nanjing 211111, China.
He is now with School of Electronic Engineering and Computer Science, Queen Mary University of London, London E1 4NS, U.K. Email: \{cunhuapan\}@seu.edu.cn. H.Zhu, Q.Z.Ahmed and J.Wang are with School of Engineering and Digital Arts, University of Kent, Canterbury, CT2 7NT,  United Kingdom. Email: \{H.Zhu,Q.Ahmed,J.Z.Wang\}@kent.ac.uk}}}
\maketitle

\begin{abstract}
Device-to-device (D2D) assisted offloading heavily depends on the participation of human users. The content preference and sharing willingness of human users are two crucial factors in the D2D assisted offloading. In this paper, with consideration of these two factors, the optimal content pushing strategy is investigated by formulating an optimization problem to maximize the offloading gain measured by the offloaded traffic. Users are placed into groups according to their content preferences, and share content with intergroup and intragroup users at different sharing probabilities. Although the optimization problem is nonconvex, the closed-form optimal solution for a special case is obtained, when the sharing probabilities for intergroup and intragroup users are the same. Furthermore, an alternative group optimization (AGO) algorithm is proposed to solve the general case of the optimization problem. Finally, simulation results are provided to demonstrate the offloading performance achieved by the optimal pushing strategy for the special case and AGO algorithm. An interesting conclusion drawn is that the group with the largest number of interested users is not necessarily given the highest pushing probability. It is more important to give high pushing probability to users with high sharing willingness.
\end{abstract}

\begin{IEEEkeywords}
Content Offloading, Device-to-Device Communications, Sharing Willingness, Content Preference
\end{IEEEkeywords}

\newpage
\section{Introduction}

Mobile communications have been developed extremely fast \cite{Zhu2009,Zhu2012,zhu2012radio,zhu2011performance,wang2012distributed,pan2016pricing,pan2015totally}.
The explosion in cellular traffic has instilled a significant strain on the current network infrastructure \cite{index2015global}. A promising solution is to offload traffic via device-to-device (D2D) communications \cite{rebecchi2015data,chen2015energy}, called D2D assisted offloading, by exploiting the caching ability of the user equipment (UE) \cite{zhang2015buffer,7544526}.
It is shown in \cite{Bastug2014} that the peak-time traffic can be substantially reduced by proactively caching the contents in UEs or BS at off-peak time. Moreover, D2D caching network with spacial reuse was proved to have great potential of improving the system throughput \cite{MJi2016}.
Specifically, in case multiple UEs request the same content from a base station (BS), instead of serving the multiple UEs individually with the requested content, the BS first pushes the content to some of these UEs. For other UEs requesting the content, if the pushed UEs are in the proximity, they can then get it via D2D communications. Otherwise, the UEs receive the requested content from the BS. By exploiting D2D assisted offloading in content dissemination, the cellular traffic load can be relieved from the cellular network infrastructure\cite{sciancalepore2016offloading,jiang2016maximized}.

The D2D assisted content dissemination process is divided into two stages, content pushing for pushing content from BS to UEs and content transmission for disseminating contents from a UE to another UE via D2D communications. It can be seen that human users are directly involved in D2D assisted content dissemination, which leads to an inevitable impact of the human behavior on the quality of service (QoS) \cite{7435232,7509388,7539688}. This situation is different from the situation in traditional network where QoS can be full controlled by the BS. Especially, content preference and sharing willingness of the human users are two crucial factors in the two stages of content dissemination.

The content preference features the different desires of human users for the same content. In this paper, content preference is modeled as the probability that a content is wanted by a UE. In the pushing stage, given UEs' content preferences, the BS will initially select some UEs to push the content. If one of the selected UEs is not interested, the UE will refuse the pushing request. Several approaches have been adopted to explain content preference\cite{7384466,7244350,li2014multiple,6928487}. A common approach is that a content consists of keywords and associated weights \cite{li2014multiple,6928487}. The weights help determine the importance of the keywords in the content. Then UEs' different preferences on the keywords will lead to different content preferences.

The sharing willingness of UEs directly affects the success of establishing D2D links, and it is a key factor in the content transmission stage. In this paper, the sharing willingness is represented by the probability that a UE will share the content via D2D communications. In current literature, the sharing willingness can be either individually based \cite{zhao2015social,li2016end,wiese2011you} or group based \cite{shaikh2009group,li2011impact}. The investigation in \cite{wiese2011you} showed that sharing willingness depends on the closeness of the social relationship between UEs. However, the complexity of tracking its social relationship with each UE is extremely high. Therefore, in \cite{shaikh2009group} sharing willingness was estimated in a group manner. Since intragroup UEs have similar human behavior, \cite{li2011impact} considered that UEs were more willing to share contents with intragroup UEs as compared to intergroup UEs.

Even though researchers have revealed that content preference and sharing willingness both affect the content dissemination process, their joint impact has not been fully investigated in the current offloading schemes.
Some offloading approaches even did not consider these two factors, such as \cite{al2014optimal, YWuEnergy2016}, where assumptions were made that all UEs in cell requested the same content and shared it altruistically. The content preference was considered in \cite{li2014multiple} and \cite{6928487}, where UEs were assumed unselfish. While in \cite{wang2015social}, only the sharing behaviors of UEs were considered for designing offloading approach, but UEs' content preferences were not addressed. Similarly, in \cite{guo2015cooperative}, although the different content preferences were considered, the sharing behaviors of UEs were simplified as only allowing sharing among a group of UEs.

Since content preference and sharing willingness of human users are all involved in D2D assisted content dissemination, it is essential to address the two factors in designing the pushing strategy to maximize the traffic offloaded via D2D communications. For instance, if the content is pushed to the interested UEs who are not willing to share, BS still has to serve the download for other interested UEs via cellular link. On the contrary, if the content is only pushed to the UEs who are willing to share but not interested in it, all the pushing requests will be refused. The interested UEs still need to be served by the BS via the cellular link. In both cases, there is no traffic offloaded via D2D communications. Therefore, current content dissemination strategies developed unilaterally either only based on content preference or sharing willingness cannot be directly applied to D2D assisted offloading.
Furthermore, the combination of the two factors further complicates the pushing strategy design.
The formulated optimization problem can be nonconvex, which cannot be solved by the interior-point methods in polynomial-time \cite{boyd2004convex}.

In this paper, content dissemination for D2D assisted offloading is designed and evaluated under the consideration of the impact of human behaviors. Especially, we investigate the pushing probabilities for different UEs to maximize the D2D offloaded traffic with the consideration of their content preferences and sharing willingness. In this paper, UEs are classified into groups according to their content preferences, and each group shares the content with inter- and intragroup UEs with different sharing probabilities. For the proposed two stages of content dissemination process, in the pushing stage, BS pushes the content to some UEs according to a pushing probability. However, the pushing will only be accepted by the UEs with interests. In the transmission stage, the probability of content sharing via D2D transmission is related to the sharing willingness of the pushed UEs in each group. Based on this content dissemination process, the pushing probability for UEs in each group is optimized in order to maximize the traffic offloaded via D2D communications. The main contributions of this paper are outlined as follows.
\begin{itemize}
\item  We define the system offloading gain as the offloaded traffic in a unit area, which is derived as a function of the pushing probability in each group. In this paper, the main aim is to maximize the system offloading gain, which is formulated to be an optimization problem, and an alternative group optimization (AGO) algorithm is proposed to solve it. Although the formulated optimization problem is nonconvex, the proposed AGO algorithm has polynomial complexity in terms of the UE group number.
\item We have derived a closed-form optimal solution by the Karush-Kuhn-Tucker (K.K.T) conditions for a special case, where UEs share contents with intergroup and intragroup UEs at the same sharing probability. It should be noted that the optimization problem is still nonconvex for this special case.
In this case, the optimal pushing strategy indicates that groups with high sharing probabilities always need to offload the large portion of traffic. The group with largest request density is not necessarily given the highest pushing probability, and its pushing probability is affected by the sharing behavior of other groups.
\item Simulation results are provided to show the offloading performance achieved by the optimal pushing strategy in the special case, where the impacts of content preference and sharing willingness are investigated. The proposed AGO algorithm is also simulated and the converged result is shown to be near the global optimum.
\end{itemize}

The rest of this paper is organized as follows. The system model is described in Section II, and the optimization problem is formulated in Section III. A closed-form optimal solution for a special case is theoretically derived in Section IV and the algorithm for the general case is introduced in Section V. Finally, we present the simulation results in Section VI, and conclude our work in Section VII.


\section{System Model}

\begin{figure}
\center
\includegraphics[width=0.5\textwidth,angle=0]{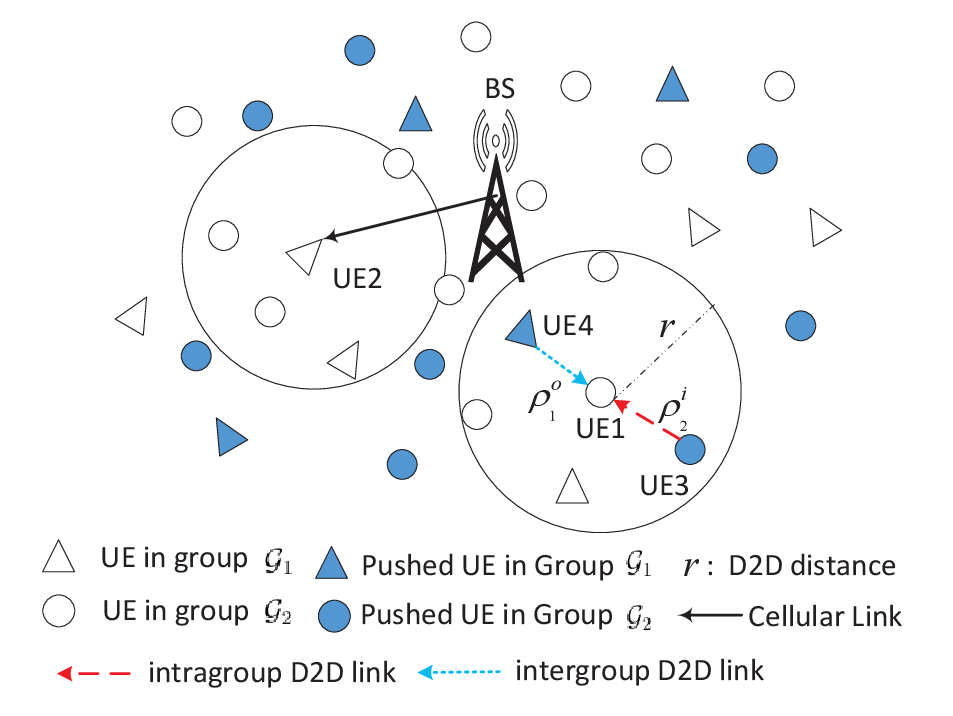}
\caption{ An example of D2D assisted offloading with $M=2$, where UE4 and UE3 provide D2D link to UE 1 with intergroup sharing probability $\rho_1^o$ and intragroup sharing probability $\rho_2^i$, respectively.}
\label{blo}
\end{figure}

In our system, the proposed content offloading scheme focuses on how to disseminate a certain piece of content (hereafter referred to as the reference content) through the cellular network, where UEs can share the cached contents via D2D communications.
As shown in Fig. \ref{blo}, the D2D transmission range is represented by radius $r$.
For content dissemination, BS first pushes the reference content to a subset of UEs denoted by the shaded circles and triangles.
Then the pushed UEs could transmit the downloaded content to other interested UEs within $r$ via D2D links.
The D2D links conducted by the pushed UEs are assumed to be all scheduled by BS.
Since all UEs receive the same content from the pushed UEs, it is assumed no co-channel interference in D2D transmissions\cite{Energy2016}.\footnotemark[1]
\footnotetext[1]{
The transmission for the pushed UEs can be modeled as a practical multicast distributed antenna system in \cite{Energy2016}. Since there is one common content for transmission, the signals from all the “pushed UEs” are regarded as useful signals at the “other UEs”. Each UE only selects its serving candidates within the D2D distance $r$.}

We assume that UEs are classified into $M$ disjoint groups according to their content preferences. The set of group index is $\mathcal{M}=\{1,2, \cdots ,M\}$. The group $m$ is denoted by $\mathcal{G}_m$.
For the given reference content, we define $w_m$ as the request probability that a UE in group $\mathcal{G}_m$ requests the content, where $0 \leq w_m \leq 1 ( m \in \mathcal{M})$. In another word, $w_m$ represents the content preference of UEs in group $\mathcal{G}_m$ for the reference content.
The value of $w_m$ for each group can be determined either by the keywords feature extraction method in \cite{li2014multiple} \cite{6928487} or by the machine learning method in \cite{chen2016echo} according to the UEs' download history.
In this paper, request probability and content preference will be used alternatively.

Another key factor is the sharing willingness of UEs to share their cached content via D2D communications, which is represented by the sharing probability.
It is observed that UEs who share common preferences are more likely to have similar personality and character \cite{lewis2012social}.
Therefore, we logically assume that the UEs in the same group also have the homogeneous sharing probability.
Here, two types of sharing probabilities are considered. We define $\rho_m^i$ as the intragroup sharing probability in group $\mathcal{G}_m$, under which a UE in group $\mathcal{G}_m$ will exchange content with another UE in the same group. $\rho_m^o$ denotes the intergroup sharing probability under which a UE in $\mathcal{G}_m$ can exchange its content with another UE in another group. Normally, it holds that $\rho_m^i \geq \rho_m^o$, due to the fact that UEs prefer to share content with those having similar preferences and social behaviors \cite{li2011impact}.

The location of the UEs in each group is modeled as a Poisson Point Process (P.P.P) \cite{chiu2013stochastic} and is independent from the other group. In P.P.P, the number of UEs in a bounded area is a Poisson random variable with a constant density. The density reflects the average number of UEs in a unit area.
 The UEs' density in group $\mathcal{G}_m$ is denoted by $\lambda_m, m \in \mathcal{M}$. Under this assumption, for a UE, it is possible to have UEs from other group in proximity.

As illustrated in Fig. \ref{blo}, according to different content preferences, UEs are divided into two groups, i.e., $\mathcal{G}_1$ and $\mathcal{G}_2$, which are represented by triangles and circles. Some UEs are able to get content from the pushed UEs of same or different groups in the proximity. For example, UE 1 gets the content from UE 3 with intragroup probability $\rho_2^i$, and from UE 4 with intergroup probability $\rho_1^o$ via D2D communications. All the nearby UEs of UE 2 do not have the content, so the BS will assist in downloading. In order to clearly describe the dissemination process we have divided it into two stages; namely, content pushing and transmission. Details of each stage are available in the following section.

\subsubsection{Content Pushing}

In this stage, the BS pushes the reference content to some UEs which are selected from all the $M$ groups. The probability of UEs in each group that will be selected for pushing will be optimized in this work. It is assumed that the selection of UEs is done randomly by the BS for the sake of fairness. For example, the shaded circles in Fig. \ref{blo} are randomly chosen from all the circles.
In previous work \cite{al2014optimal, YWuEnergy2016, li2014multiple, 6928487, zhao2015social, wang2015social, guo2015cooperative}, it was assumed that UEs who have received a pushing request will always accept it. However, UEs will accept the pushing only when they are interested in the pushed content. Otherwise, they will simply ignore it and refuse the pushing. Therefore, there will be four kinds of UEs at this stage as shown in Table \ref{table state}.

\begin{table}[H]
\centering
\caption{UE types in content pushing stage}
\label{table state}
\begin{tabular}{c|c|c|c}\hline
UE Type & pushing request ? & interested ? & UE behavior \\
\hline
UE-A & Received & Yes &accept and download content \\
\hline
UE-R & Received & No &refuse or ignore pushing request \\
\hline
UE-T & Not received & Yes & ask for content transmission \\
\hline
UE-N & Not received & No & irrelevant with dissemination\\ \hline
\end{tabular}
\end{table}

In Table \ref{table state}, when UEs have received the content pushing request, those who accept the request and download the content from BS are referred to as “UE-A”s,
while those who refuse the pushing request are called as “UE-R”s.
For the other UEs who do not receive pushing request, if they are interested in the reference content, such as UE 1 in Fig. 1, they will ask for content transmission.
These UEs are named as the “UE-T”s.
Finally, UEs who are not involved in the content dissemination process are represented as “UE-N”s.

Let $c_m$ be the pushing probability for UEs in group $\mathcal{G}_m$. $c_m$ also represents the probability that a UE in group $\mathcal{G}_m$ will receive the pushing from the BS. For a UE being a UE-A, there are two conditions. One is that the UE wants this content, and the other is that a pushing has been received. Given the UEs density $\lambda_m$ and content  request probability $w_m$, under the P.P.P model, we can calculate the density of UE-As in group $\mathcal{G}_m$ denoted by $l_m$ as
\begin{equation}
l_m = \lambda_m w_m c_m.
\end{equation}

Similarly, the density of UE-Ts in group $\mathcal{G}_m$ denoted as $n_{m}$ is obtained as
\begin{equation} \label{nm}
n_{m} = \lambda_m w_m (1-c_m).
\end{equation}
The density of UE-Rs and UE-Ns can also be calculated in the same way, which are ignored for brevity since they are not involved in the upcoming analysis.

\subsubsection{Content Transmission}

Since the UE-Ts did not get the reference content in the content pushing stage, they will make transmission requests to download it from the BS or the nearby UE-As. Let $\mathbb{P}_m$ denote the probability that a UE-T in $\mathcal{G}_m$ can download the content via D2D links, which means there is at least a nearby UE-A will share it via D2D transmission. $\mathbb{P}_m$ is also called as the D2D success probability and will be derived in the rest of this section.

For the P.P.P distribution, the number of UEs is a Poisson random variable with density $\lambda$. Therefore, the probability that there are $n$ UEs in area $A$ is calculated as
\begin{equation}\label{po}
P(n,A)= \frac{(\lambda A)^n}{n!}\textrm{exp}(-\lambda A).
\end{equation}
Since the content holders are composed of the intragroup UE-As and intergroup UE-As due to the social sharing willingness. To calculate the density of content holders, we need to get the density of intragroup UE-As and intergroup UE-As, respectively.
For a UE-T in $\mathcal{G}_m$, let $L_m$ denote the density of intragroup UE-As willing to share the content via D2D transmissions. It is obtained as
\begin{equation}
L_m=\rho_m^i l_m=\lambda_m w_m \rho_m^i c_m.
\end{equation}
The density of intergroup UE-As from other groups with the willingness to share the content via D2D transmissions is denoted by $O_{m}$, which is given by
\begin{equation}\label{om}
 O_{m}=\sum_{k \neq m}\rho_k^o l_k= \sum_{k \neq m}\lambda_k w_k  \rho_k^o c_k.
\end{equation}

Based on (\ref{po})-(\ref{om}), the probability that no UE within D2D transmission distance $r$ will transmit this content to a UE-T in $\mathcal{G}_m$ is calculated as
\begin{equation}
P(0,\pi r^2) = \textrm{exp}(-\pi r^2(L_m+O_m)),
\end{equation}
and the success probability $\mathbb{P}_m$ that a UE-T in $\mathcal{G}_m$ can download the content by D2D offloading is obtained as
\begin{equation} \label{pm}
\mathbb{P}_m =1-P(0,\pi r^2)=1-\textrm{exp}(-\pi r^2(L_m+O_m)).
\end{equation}

\section{Problem Formulation}
With the system model in place, we will now characterize the quality of content service for UEs.
Given the UE-T density $n_m$ and the D2D success probability $\mathbb{P}_m$ for UE-Ts, the offloading gain $G_m$ of group $\mathcal{G}_m$ is defined as
\begin{equation}
G_m = n_{m}\mathbb{P}_m.
\end{equation}
$G_m$ can be regarded as the expected offloaded traffic of group $\mathcal{G}_m$ in a unit area, which is similar to the measure of offloading gain used in \cite{chen2016cache}.

The system offloading gain $G$ is the performance measure for a D2D assisted offloading network. $G$ is defined as the sum of offloading gain over all the groups.
\begin{equation}\label{G}
G=  \sum \limits_{m \in \mathcal{M}} G_m = \sum \limits_{m \in \mathcal{M}} n_{m}\mathbb{P}_m.
\end{equation}
It can be seen from (\ref{G}) that $G$ represents the successfully offloaded content copies from all the $M$ groups, therefore, reflecting the offloading ability of the system. Substituting $n_{m}$ and $\mathbb{P}_m$ in (\ref{G}) with (\ref{nm}) and (\ref{pm}), we have the following equivalent expression,
\begin{equation}\label{Gc}
G = \sum \limits_{m \in \mathcal{M}} \lambda_m w_m  (1- c_m)
( 1 - \textrm{exp}(-B \lambda_m w_m \rho_m^i c_m
 -B \sum_{k \neq m} \lambda_k w_k \rho_k^o c_k) ).
\end{equation}
where $B$ represents the D2D cooperation area, i.e., $B=\pi r^2$.

Given the request probability, $w_m$, user density, $\lambda_m$, intergroup sharing probability, $\rho_m^i$, and the intragroup sharing probability, $\rho_m^o$, of each group, $G$ is then determined by the pushing probability $c_m$ in all the $M$ groups. In order to show the relationship between $G$ and $c_m$, we consider two extreme cases; namely, \textit{all pushing case} and \textit{all request case}.
In \textit{all pushing case}, $c_m = 1$ for all $m$ ($m \in \mathcal{M}$). Then $n_{m}=0$ for all the $M$ groups. While, in \textit{all request case}, $c_m = 0$ for all $m$. Then $\mathbb{P}_m=0$ for all $m$. In both cases, $G=0$. Therefore, we aim to find out the optimal value of pushing probability $c_m$ for each group $\mathcal{G}_m$ to maximize $G$.

As our objective is to maximize the offloading gain, the optimization problem is formulated as
\begin{subequations}
\begin{align}
 \mathcal{P}1:\mathop{\max }_{\bm{c}}
                  &\quad  G(\bm{c}) =\sum \limits_{m \in \mathcal{M}} n_{m}\mathbb{P}_m,    \label{opt1}   \\
\textrm{s.t.} &\quad 0\leq c_m \leq 1 , \textrm{ for all } m \in  \mathcal{M}. \label{st}
\end{align}
\end{subequations}
In problem $\mathcal{P}1$, vector $\bm{c}$ in (\ref{opt1}) is described by $\bm{c} = [c_1,c_2,\cdots,c_M]$, and represents the pushing strategy of the system.
The constraint (\ref{st}) ensures that $c_m, \forall m \in \mathcal{M}$ is a valid probability. For the groups with zero request probabilities, the optimal pushing probabilities are zeros, since all the pushing requests will be refused. Therefore, in the following solution we assume that all the $M$ groups have positive request probabilities, i.e., $w_m>0$ for all $m$.

\begin{figure}
\centering
\includegraphics[width=0.75\textwidth]{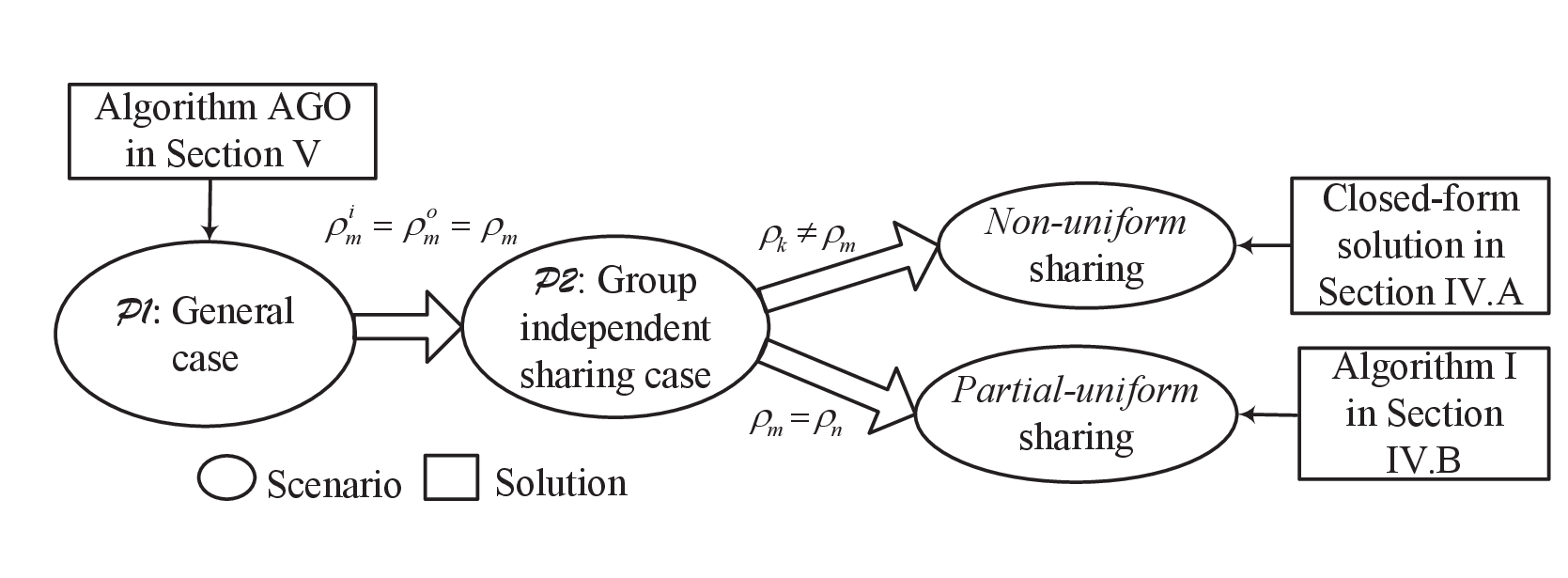}
\caption{The considered scenarios and the corresponding solutions.}
\label{figSS}
\end{figure}

As shown in Fig. \ref{figSS}, the solutions of  $\mathcal{P}1$ are presented for different scenarios in the following section IV and V.
First of all, it is easily verified that problem $ \mathcal{P}1$ is nonconvex.
Therefore, AGO algorithm with polynomial complexity is proposed in Section V to solve it.
However, the optimal analytical solution can be obtained for a special case, when intragroup sharing probability equals intergroup sharing probability, i.e., $\rho_m^i=\rho_m^o=\rho_m$ for any $m$.
We refer to this case as the \textit{group independent sharing case}.
Moreover, based on the sharing probability distribution, the \textit{group independent sharing case} is divided into a \textit{non-uniform} sharing scenario and a \textit{partial-uniform} sharing scenario.
In \textit{non-uniform} sharing scenario, the sharing probabilities are different for different groups, i.e., $\rho_k \neq \rho_m$, if $m \neq k$.
In \textit{partial-uniform} sharing scenario, a part of groups have the same sharing probability, i.e., there exists $\rho_{m}= \rho_{n}, n \neq m$.
The uniform scenario where every group has the same sharing probability is a special case of \textit{partial-uniform} sharing scenario, and thus it is not listed separately.
The reason for this classification is that the optimal pushing strategy for \textit{partial-uniform} sharing scenario is not unique.
However, a special case of the alternative optimal solutions will be investigated in Algorithm 1 of Section IV B.

In addition, this classification can be applied to corresponding practical scenarios.
A residential area is an example for \textit{non-uniform} sharing scenario, where the residents are of all ages.
The reason is that people in different ages normally have different content preferences and sharing behaviors.
A campus network is a practical scenario for \textit{partial-uniform} sharing, where students prefer different contents but all are willing to share.

\section{Solution Analysis for Group Independent Sharing Case}

\newtheorem{proposition}{\textbf{Proposition}}[section]
\newtheorem{theorem}{\textbf{Theorem}}[section]
\newtheorem{lemma}{\textbf{Lemma}}[section]
\newtheorem{corollary}{\textbf{Corollary}}[section]

In the \textit{group independent sharing case}, since the intragroup and intergroup sharing probability are the same, problem $\mathcal{P}1$ is simplified as
\begin{subequations}
\begin{align}
 \mathcal{P}2:\mathop{\max }_{\bm{c}}
                  &\quad   G_{f}(\bm{c})= \left(\sum \limits_{m \in \mathcal{M}} t_m  (1- c_m)\right)
             \left( 1 - \textrm{exp}{(-B\sum\limits_{k \in \mathcal{M}} t_k \rho_k c_k  ) }  \right),  \\
\textrm{s.t.} &\quad 0\leq c_m \leq 1 ,\quad m \in  \mathcal{M}.
\end{align}
\end{subequations}
where $t_m=\lambda_m w_m$ is the requested density and represents the average number of UEs from group $\mathcal{G}_m$ in unit area requesting this content.

\subsection{Non-uniform Sharing}

In this section, we discuss the solution for the \textit{non-uniform} sharing scenario.
Although Problem $\mathcal{P}2$ is simplified, it is still nonconvex.
However, by the following proof line, we can obtain the optimal pushing strategy in a closed form.
First, a special structure of the optimal solution is revealed by Lemma \ref{Theorem1}.
Second, this special structure is shown to be associated with the order of sharing probabilities in Lemma \ref{Theorem2}.
Then, by applying the special structure of the optimal pushing strategy in K.K.T conditions, one case of the optimal solutions is given in Theorem \ref{Theorem3}.
Finally, the general closed-form expression of the optimal pushing strategy is summarized in Theorem \ref{Theorem4}.

\begin{lemma}\label{Theorem1}
In the optimal pushing strategy $\bm{c}^{\ast} =[c_1^{\ast}, c_2^{\ast}, \cdots, c_M^{\ast}]$, at most one group has the optimal pushing probability lying in the range $0 < c_i^{\ast} < 1$, and for all the other groups, i.e.,  $j \neq i$,  $c_j^{\ast} = 0$ or $c_j ^{\ast}= 1$.
\end{lemma}
\textit{Proof}: Refer to the Appendix A for the detailed proof.
$\hfill\blacksquare$

To find the optimal solution, we invoke Lemma \ref{Theorem1} and investigate the relationship between the pushing probability, $c_m$, and sharing probability, $\rho_m$, in the following lemma.​
\begin{lemma}\label{Theorem2}
If the optimal pushing probability for group $\mathcal{G}_m$ is between 0 and 1, i.e., $0<c_m^{\ast} <1$, then for any group $\mathcal{G}_{i}$ with $\rho_i < \rho_m$, the optimal pushing probability is $c_{i}^{\ast} =0$, and for any group $\mathcal{G}_{j}$ with $\rho_j > \rho_m$, the optimal pushing probability is $c_{j}^{\ast} =1$.
\end{lemma}
\textit{Proof}: Refer to the Appendix B for the detailed proof.
$\hfill\blacksquare$

According to Lemma \ref{Theorem2}, when the groups are sorted in the ascending order of sharing probabilities, i.e., $\rho_1 < \rho_2 <\cdots < \rho_M$, there exists a special group $\mathcal{G}_{m}$ called the ``\textit{watershed}" with respect to the all pushing groups ($c_j^{\ast}=1$) and non-pushing groups ($c_i^{\ast}=0$). If the ``\textit{watershed}"  group $\mathcal{G}_{m}$ is determined, the optimal pushing strategy for this D2D offloading system is obtained, which can be written as $\bm{c}^{\ast} =[\underbrace{0, \cdots, 0}_{m-1},c_m^{\ast},\underbrace{1, \cdots, 1}_{M-m}]$, where $(m-1)$ zeros represents the non-pushing group while $(M-m)$ ones are the all pushing groups. ​

Furthermore, proof of Lemma \ref{Theorem2} leads to the following two corollaries.
\begin{corollary}\label{Cor2}
If $c_m^{\ast} =0$, for group $\mathcal{G}_m$, $c_{i}^{\ast}= 0$ for group $\mathcal{G}_{i}$ where $\rho_i < \rho_m$. If $c_m^{\ast} =1$, $c_{j}^{\ast}= 1$ for group $\mathcal{G}_{j}$ where $\rho_j > \rho_m$.
\end{corollary}
Similar as Lemma \ref{Theorem2}, Corollary \ref{Cor2} can also be proved by K.K.T conditions. Details are omitted for brevity.

The optimal pushing probability for the ``\textit{watershed}"  group is determined according to the following Corollary \ref{Corcm}.
\begin{corollary}\label{Corcm}
Assuming that $M$ groups are sorted in the ascending order of sharing probabilities, i.e., $\rho_1 < \rho_2 <\cdots < \rho_M$. For group $\mathcal{G}_m$, if the optimal pushing probability $0<c_m^{\ast}<1$, then
\begin{equation}
c_m^{\ast}= \frac{1}{B\rho_m t_m}
\left( B\rho_m\sum\limits_{i=1}^{m} t_i +1 -\mathcal{W}(\textrm{exp}(B\sum\limits_{j=m+1}^{M} \rho_j t_j + B\rho_m\sum\limits_{i=1}^{m} t_i +1 )) \right),  \label{cm}
\end{equation}
where $\mathcal{W}$ is the Lambert-W function \cite{weisstein2002lambert}.

\end{corollary}
\textit{Proof}: Refer to the Appendix C for the detailed proof.
$\hfill\blacksquare$

However, the following key problem is to find the ``\textit{watershed}"  group.
To solve this problem, Theorem \ref{Theorem3} is introduced to show the necessary and sufficient conditions of ``\textit{watershed}"  group.

\begin{theorem}\label{Theorem3}
Assuming that $M$ groups are sorted in the increasing order of $\rho_m$, i.e., $\rho_1 < \rho_2 <\cdots < \rho_M$, the optimal solution of problem $\mathcal{P}2$ is $\bm{c}^{\ast} =[\underbrace{0, \cdots, 0}_{m-1},c_m^{\ast},\underbrace{1, \cdots, 1}_{M-m}]$, where $c_m^{\ast}$ is given by (\ref{cm}), \textit{ if and only if} the following two inequalities
\begin{numcases}{}
1+B \rho_m \sum\limits_{i=1}^{m} t_i > \textrm{exp}(B\sum\limits_{j=1+m}^{M} t_j\rho_j), \label {1Condition}\\
1+B \rho_m \sum\limits_{i=1}^{m-1} t_i < \textrm{exp}(B\sum\limits_{j=m}^{M} t_j\rho_j).  \label {0Condition}
\end{numcases}
hold at the same time.
\end{theorem}
\textit{Proof}: Refer to the Appendix D for the detailed proof.
$\hfill\blacksquare$

Theorem \ref{Theorem3} shows how request density and sharing probability jointly impact the optimal pushing strategy. Furthermore, it is worth noting that the conditions in Theorem \ref{Theorem3} guarantee that (\ref{cm}) is always feasible. Finally, the uniqueness proved by Lemma \ref{proUni} is consistent with Lemma \ref{Theorem1}, which shows that the pushing strategy given by Theorem \ref{Theorem3} is exclusive.
For simplicity, we define two functions as follows,
\begin{eqnarray}
f^1_m(t_m, \rho_m) = 1+B \rho_m \sum\limits_{i=1}^{m} t_i - \textrm{exp}\left(B\sum\limits_{j=1+m}^{M} t_j\rho_j\right), \label{f1m}\\
f^0_m(t_m, \rho_m) = \textrm{exp}\left(B\sum\limits_{j=m}^{M} t_j\rho_j\right)-B \rho_m \sum\limits_{i=1}^{m-1} t_i -1.\label{f0m}
\end{eqnarray}
From Theorem \ref{Theorem3}, we can infer the following two corollaries.

\begin{corollary}\label{C0}
For group $\mathcal{G}_m$, if $f^1_m(t_m, \rho_m)\leq 0$, then $c_m^{\ast}=0$;
\end{corollary}
\begin{corollary}\label{C1}
For group $\mathcal{G}_m$, if $f^0_m(t_m, \rho_m) \leq 0$, then $c_m^{\ast}=1$.
\end{corollary}
The proofs of Corollary \ref{C0} and Corollary \ref{C1} are also carried out by contradiction, which is similar to the ``\textit{if}" proof part in Theorem \ref{Theorem3}. Details are omitted for brevity.
Based on the foregoing analysis, at the \textit{non-uniform} sharing scenario, a closed-form optimal solution of the nonconvex problem $\mathcal{P}2$ is summarized in the following theorem.
\begin{theorem}\label{Theorem4}
Assuming that $M$ groups are sorted in the order $\rho_1 < \rho_2 <\cdots < \rho_M$, the optimal solution of problem $\mathcal{P}2$ is given as
\begin{equation}
\bm{c}^{\ast} =\left\{\begin{matrix}
[\underbrace{0,\cdots,0}_{m},\underbrace{1,\cdots,1}_{M-m}], \textrm{for all $m$}
\begin{cases}
 f^1_m(t_m, \rho_m)\leq 0, \\
 f^0_{m+1}(t_{m+1}, \rho_{m+1})\leq 0. \\
\end{cases}\\
[\underbrace{0,\cdots,0}_{m-1},c_m,\underbrace{1,\cdots,1}_{M-m}], \textrm{there exists an  $m$}
\begin{cases}
 f^1_m(t_m, \rho_m) > 0, \\
 f^0_m(t_m, \rho_m) > 0.\\
\end{cases}
\end{matrix}\right.
\end{equation}
where $c_m$ is given by (\ref{cm}).
\end{theorem}
\textit{Proof}: The first case is readily obtained by Corollary \ref{C0} and Corollary \ref{C1}, and the second case is obtained by Theorem \ref{Theorem3}. Thus, Theorem \ref{Theorem4} is proved.
$\hfill\blacksquare$

\begin{algorithm}
\caption{: A special case of optimal pushing strategy for \textit{partial-uniform} sharing scenario}  \label{alg1}
\begin{algorithmic}
\STATE
\textit{Step 1:} Sort the $M$ groups in the  ascending order of $\rho_m$, i.e.,  $\rho_1 <\cdots <\rho_{k_1}= \cdots =\rho_{k_n}<\cdots< \rho_M$.
\STATE
\textit{Step 2:} Define group 0 as $t_0 = \sum\limits_{k=k_1}^{k_n} t_k, c_0 =\frac{1}{t_0}\sum\limits_{k=k_1}^{k_n} t_kc_k$.  Substitute group 0 for the $n$ groups, i.e., $\mathcal{G}_{k_1}$, $\cdots$ $\mathcal{G}_{k_n}$.
\STATE
\textit{Step 3:} Calculate the optimal pushing strategy $\bm{c}^{\ast} =[c_1^{\ast},\cdots,c_0^{\ast},\cdots,c_M^{\ast}]$ for the $M-n+1$ groups according to Theorem \ref{Theorem4}.
\STATE
\textit{Step 4:} Set the pushing probabilities of the replaced $n$ groups to be the same as $c_0^{\ast}$.
\end{algorithmic}
\end{algorithm}

\subsection {Partial-uniform Sharing}
In the \textit{partial-uniform} sharing scenario, we will prove that the optimal pushing strategy is not unique, and a special case of the alternative optimal pushing strategies will be given in this section.
\begin{theorem}\label{Theorem5}
If $n$ groups have the same sharing probability, where $2 \leq n \leq M$, the optimal pushing probabilities of these $n$ groups are not unique.
\end{theorem}
\textit{Proof}: Refer to the Appendix F for the detailed proof.
$\hfill\blacksquare$

Based on the proof of Theorem \ref{Theorem5}, we propose an algorithm to find a special case of the alternative optimal pushing strategies for the \textit{partial-uniform} sharing scenario, which is described in Algorithm \ref{alg1}.
It is worth pointing that the pushing strategy obtained in Algorithm \ref{alg1} is the optimal solution, and the obtained offloading gain achieves the maximum value.
It is shown in the proof of  Theorem \ref{Theorem5} that the variable substitution (group 0) in \textit{Step 2} transforms the \textit{partial-uniform} sharing scenario into an equivalent  \textit{non-uniform} sharing scenario.
In addition, according to Theorem \ref{Theorem4}, the pushing strategy for the equivalent  \textit{non-uniform} sharing scenario in \textit{Step 3} is the global optimal.
Therefore, the solution in \textit{Step 4} is an optimal solution, although it is not unique.

\subsection {Impact of System Parameters}
In this section, a brief discussion is conducted to show the insights given by the optimal pushing strategy.
According to the above solution analysis, the optimal solution of \textit{partial-uniform} sharing scenario is obtained by transforming it to the equivalent \textit{non-uniform} sharing scenario.
Therefore, the following discussion is based on the analytical solution obtained in the \textit{non-uniform} sharing scenario.

\begin{figure}
\centering
\includegraphics[width=0.55\textwidth]{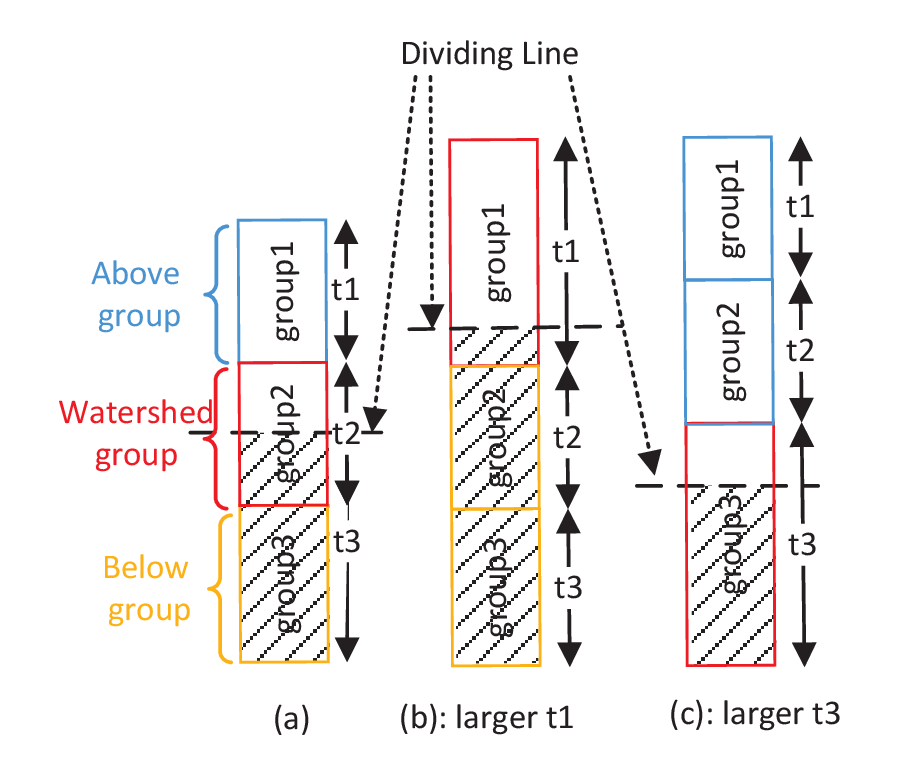}
\caption{An example of the optimal pushing strategy, where the shaded ratio of each group represents its pushing probability.
The blue blocks represent the above groups, red blocks represent the watershed groups and the yellow ones represent the below groups.
Compared with case (a), case (b) has a larger $t_1$, and it leads to a higher ``dividing line".
While case (c) has a larger $t_3$ , and it leads to a lower ``dividing line". }
\label{figlull}
\end{figure}

An example of the pushing strategy is shown in Fig. \ref{figlull}, where 3 groups are sorted in the order $\rho_1 < \rho_2 <\rho_3$.
The length of each group is their request density, and the shaded proportion in each group represents its pushing probability.
For example, in the case (a) of Fig. \ref{figlull}, group 3 is all in shadow as it has the 100\% pushing probability.
Group 2 is partial shaded due to that the pushing probability is between 0 and 1.
There is no shadow part in group 1 resulting from no pushing.
Therefore, there is a ``dividing line" which is responsible to define three types of groups in the optimal pushing strategy;
The ``\textit{watershed}" group with ``dividing line", the ``\textit{below}" groups with 100\% pushing and the ``\textit{above}" groups with 0\% pushing.
These three types of groups play different roles in determining the ``dividing line", which are explained in details separately.

\subsubsection{Group $\mathcal{G}_{m}$ is a ``\textit{watershed}" group}
\begin{proposition}\label{Prop1}
When the request density $t_m \in (0, \infty)$ increases, the ``dividing line" goes down, but it is still located in group $\mathcal{G}_{m}$.
\end{proposition}
\textit{Proof}: By taking derivatives of $f^1_m(t_m, \rho_m)$ with respect to (w.r.t) $t_m$ and $f^0_m(t_m, \rho_m)$ w.r.t $t_m$ respectively, we have
\begin{equation}
\frac{\partial f^1_m(t_m, \rho_m) }{\partial t_m} = B\rho_m,
\frac{\partial f^0_m(t_m, \rho_m) }{\partial t_m} = B\rho_m\textrm{exp}\left(B\sum\limits_{j=m}^{M} t_j\rho_j\right).
\end{equation}
Therefore, the conditions that $ f^1_m(t_m, \rho_m)>0 ,f^0_m(t_m, \rho_m) >0$ are always satisfied for group $\mathcal{G}_{m}$ when $t_m$ increases.
In addition, we define function
\begin{equation} \label{gcm}
g(c_m^{\ast},t_m)=\left(B \rho_m\left({\sum\limits_{i=1}^{m}t_i}-t_m c_m^{\ast}\right) + 1\right) \textrm{exp}(-B t_m\rho_m c_m^{\ast}).
\end{equation}
From (\ref{EquCm}) in Appendix C, the following equation is obtained.
\begin{equation}\label{Eqs}
g(c_m^{\ast},t_m)= \textrm{exp}\left(\sum\limits_{j=m+1}^{M} B t_j \rho_j \right).
\end{equation}
Given the request density and the sharing probability in each group, the RHS of (\ref{Eqs}) is a positive fixed value.
By taking the derivatives of $g(c_m^{\ast},t_m)$ w.r.t $c_m^{\ast}$ and $t_m$, we have
\begin{equation}\label{r1}
\frac{\partial g(c_m^{\ast}) }{\partial c_m^{\ast}} = -B t_m\rho_m\left(B \rho_m\left({\sum\limits_{i=1}^{m}t_i}-t_m c_m^{\ast}\right) + 2\right) \textrm{exp}(-B t_m\rho_m c_m^{\ast}) < 0,
\end{equation}
\begin{equation}\label{r2}
\frac{\partial g(c_m^{\ast}) }{\partial t_m} = -B c_m^{\ast}\rho_m\left(B \rho_m\left({\sum\limits_{i=1}^{m}t_i}-t_m c_m^{\ast}\right) + 2\right) \textrm{exp}(-B t_m\rho_m c_m^{\ast})  < 0.
\end{equation}
Regarding to the results in (\ref{r1}) and (\ref{r2}), $c_m^{\ast}$ decreases when $t_m$ increases. However, group $\mathcal{G}_{m}$ is still the ``\textit{watershed}" group.
$\hfill\blacksquare$

To explain it further, as illustrated in the case (a) of Fig. \ref{figlull}, increase of $t_2$ makes the shaded ratio of group 2 decrease, but this will not formulate another group to be the ``\textit{watershed}" group. Unfortunately, the impact of sharing willingness $\rho_m$ on the pushing strategy is more complex, which will be explained in the simulation results.

\subsubsection{Group $\mathcal{G}_{m}$ is an ``\textit{above}" group}
\begin{proposition}\label{Prop2}
When the request density $t_m \in (0, \infty)$ increases, the ``dividing line" goes up.
\end{proposition}
\textit{Proof}: We assume that group $\mathcal{G}_{k}$ is the ``\textit{watershed}" group.
The following derivatives are obtained.
\begin{equation}
\frac{\partial f^1_k(t_k, \rho_k) }{\partial t_m} = B\rho_m, \frac{\partial f^0_k(t_k, \rho_k) }{\partial t_m} = -B\rho_m.
\end{equation}
Therefore, with the increase of $t_m$, $f^1_k(t_k, \rho_k)>0$ always holds for $\mathcal{G}_{k}$. Nevertheless, $f^0_m(t_m, \rho_m)$ decreases with $t_m$.

If condition $f^0_k(t_k, \rho_k) >0 $ is not satisfied, $\mathcal{G}_{k}$ is not the ``\textit{watershed}" group and $c_k^{\ast}=1$.
In this case, since $f^0_k(t_k, \rho_k)$ is a decreasing function of $k$, only the group $\mathcal{G}_{i}, \forall i <k$ has the possibility to be the next ``\textit{watershed}" group.

If condition $f^0_k(t_k, \rho_k)>0$ is satisfied, for the ``\textit{watershed}" group $\mathcal{G}_{k}$, (\ref{gcm}) is rewritten as
\begin{equation}
g(c_k^{\ast},t_m)=\left(B \rho_k\left({\sum\limits_{i =1,i\neq m}^{k}t_i}+ t_m-t_k c_k^{\ast}\right) + 1\right) \textrm{exp}(-B t_k\rho_k c_k^{\ast}).
\end{equation}
The derivative of $g(c_k^{\ast},t_m)$ w.r.t $t_m$ is obtained as
\begin{equation}
\frac{\partial g(c_k^{\ast},t_m) }{\partial t_m}=B \rho_k\textrm{exp}(-B t_k\rho_k c_k^{\ast}).
\end{equation}
Furthermore, from (\ref{EquCm}) in Appendix C, the following equation holds for $g(c_k^{\ast},t_m)$.
\begin{equation}\label{Eqg}
g(c_k^{\ast},t_m)= \textrm{exp}\left(\sum\limits_{j=k+1}^{M} B t_j \rho_j \right).
\end{equation}
Since $g(c_k^{\ast},t_m)$ is a decreasing function of $c_k^{\ast}$, we can claim that $c_k^{\ast}$ increases with $t_m$.
$\hfill\blacksquare$

Overall,  the ``dividing line" goes up when the request density of an ``\textit{above}" group increases.
For example, in Fig. \ref{figlull}, group 1 in case (b) has a larger request density $t_1$ compared with case (a). This leads to the condition that $f^0_2(t_2, \rho_2)>0$ is no longer satisfied for group 2. The ``dividing line" goes up to group 1, and it becomes the new ``\textit{watershed}"  group.

\begin{proposition}\label{Prop3}
The optimal solution $\bm{c}^{\ast}$ keeps the same with the change of $\rho_m \in (0, \rho_{k})$, as long as the constraint $\rho_m< \rho_{k}$ holds, and $\rho_{k}$ is the sharing probability of ``\textit{watershed}" group.
\end{proposition}
\textit{Proof}: As shown in (\ref{cm}) (\ref{f1m}) and (\ref{f0m}), when $\rho_m< \rho_{k}$, $\rho_m$ is not involved in the expressions of $f^1_k(t_k, \rho_k)$, $f^0_k(t_k, \rho_k)$ and $\bm{c}^{\ast}$. Therefore, the proposition is proved.
$\hfill\blacksquare$

Proposition \ref{Prop3} implies that the D2D transmission for UEs in the ``\textit{above}" group are offered by the groups with high sharing probability. This leads to the result that the pushing strategy does not change when the sharing probabilities of ``\textit{above}" groups increase.

\subsubsection{Group $\mathcal{G}_{m}$ is a ``\textit{below}" group}
\begin{proposition}\label{Prop4}
When $t_m \in (0, \infty) $ and $\rho_m\!\! \in (\rho_{m-1}, 1]$ increases, the ``dividing line" declines.
\end{proposition}
\textit{Proof}: The proof is similar with proposition \ref{Prop2}. Details are omitted for brevity. $\hfill\blacksquare$

For example, in Fig. \ref{figlull}, group 3 in case (c) has a larger $t_3$ compared with case (a). In case (c), the condition $f^1_2(t_2, \rho_2)>0$ is no longer satisfied for group 2, and group 3 becomes the ``\textit{watershed}"  group.

\section{Solution Analysis for problem $\mathcal{P}1$}

When the intragroup sharing probabilities are different with the intergroup sharing probabilities, i.e., $\rho_m^i \neq \rho_m^o$, the optimal pushing strategy to the \textit{Group Independent Case} is feasible but not an optimal solution for problem $\mathcal{P}1$.
To solve problem $\mathcal{P}1$, we propose an alternative group optimization algorithm (AGO) in this section.

For problem $\mathcal{P}1$, it is hard to find the closed-form optimal solution due to its nonconvexity.
However, this problem has a nice property to be explored as follows.
For each group $\mathcal{G}_m$, if the pushing probabilities of other groups $\{\mathcal{G}_j| j \neq m\}$ are given, the optimal pushing probability of $\mathcal{G}_m$ can be achieved.
For group $\mathcal{G}_m$, the pushing probabilities of other groups $\{\mathcal{G}_j| j \neq m\}$ are denoted by a vector $\bm{c}_{-m} = [c_1, \cdots,c_{m-1},c_{m+1},\cdots, c_M]$.
If $\bm{c}_{-m}$ is given, the offloading gain is dependent on $c_m$, i.e $G(c_m|\bm{c}_{-m})$.
In this case, the objective function (\ref{opt1}) in problem $\mathcal{P}1$ is reduced to
\begin{equation}
G_{-m}(c_m) = t_m  (1- c_m)\left (1 - \textrm{exp}(-B t_m \rho_m^i c_m - \varphi_m \right) + \sum_{k \neq m}^M Q_k\left(1 - \textrm{exp}(-B t_m \rho_m^o c_m - \Phi_k)\right) \nonumber.
\end{equation}
where
$\varphi_m =\sum_{k \neq m}^M B t_k \rho_k^o c_k,
Q_k = t_k  (1- c_k),
\Phi_k = \sum_{q \neq m,k}^M B t_q \rho_q^o c_q + B t_k \rho_k^i c_k.$

Therefore, we have the following problem formulated.
\begin{subequations}
\begin{align}
 \mathcal{P}3:\mathop{\max }_{c_m}
                  &\quad   G_{-m}(c_m),  \label{opt3}\\
\textrm{s.t.} &\quad 0\leq c_m \leq 1.
\end{align}
\end{subequations}

The second derivative of the objective function (\ref{opt3}) is derived as
\begin{eqnarray} \label{2rd}
\frac{\partial^2 G_{-m}(c_m) }{\partial c_m^2}
= - B t_m^2 \rho_m \textrm{exp}(-B t_m \rho_m^i c_m - \varphi_m) (2+ B t_m\rho_m^i (1- c_m)) \nonumber \\
- (B t_m \rho_m^o)^2 \sum_{k \neq m}^M Q_k \textrm{exp}(-B t_m \rho_m^o c_m - \Phi_k).
\end{eqnarray}
Thus, we can claim that the objective function (\ref{opt3}) is a concave function, and problem $\mathcal{P}3$ is convex. Thanks to the convexity of $ \mathcal{P}3$, the optimal pushing probability is derived in the following proposition.

\begin{proposition}\label{Theorem6}
The optimal pushing probability for group $\mathcal{G}_m$ in problem $\mathcal{P}3$ is $c_m^{\ast}=[\Gamma_m]^1_0$, where $\Gamma_m$ is the unique solution of equation $g_m(x) =0$.
\begin{equation}
g_m(x) = (1+ B t_m \rho_m^i (1-x)) \textrm{exp}({-B t_m \rho_m^i x - \varphi_m})
+ B I_m \rho_m^o\textrm{exp}(-{B t_m \rho_m^o x}) -1,
\end{equation}
where $I_m=\sum_{k \neq m}^M Q_k \textrm{exp}(- \Phi_k)$. $x=[a]^1_0$ means that if $a>1$, $x=1$, if $a<0$, $x=0$; and if $0\leq a \leq 1$, $x=a$.
\end{proposition}
\textit{Proof}: Refer to the appendix G for the detailed proof.
$\hfill\blacksquare$

\begin{algorithm}
\caption{Alternative Group Optimization  Algorithm for solving problem $\mathcal{P}1$}  \label{alg2}
\begin{algorithmic}
\REQUIRE ~~UE densities $\lambda_m$; popularities $w_m$; D2D cooperation area $B$;maximum iterations $I_{max}$ \\
\quad \quad~~  intra- and intergroup sharing probabilities $\bm{\rho}_{in}=[\rho_1^i, \cdots,\rho_M^i]$, $\bm{\rho}_{out}=[\rho_1^o, \cdots,\rho_M^o]$;\\
\STATE  \textit{Step 1:} Initialize $\bm{c}^0=[0,\cdots,0]$; iteration number $i=0$;
\STATE  \textit{Step 2:} Repeat for iteration $i=1,2, \cdots, I_{max}$
\STATE  \textit{Step 3:} Repeat for group $m=1,2, \cdots, M$    \\
\STATE \textit{Step 4:} Update $\bm{c}_{-m}^i = [c_1^{i},\cdots,c_{m-1}^{i},c_{m+1}^{i-1},\cdots,c_{M}^{i-1}]$  \\
\STATE \textit{Step 5:} Update $c_m^{i}$ according to proposition \ref{Theorem6}. \\
\STATE \textit{Step 6:} Update $\bm{c}^{i,m}= [c_1^{i},\cdots,c_{m}^{i},c_{m+1}^{i-1},\cdots, c_{M}^{i-1}]$.\\
\STATE \textit{Step 7:} Update $G^{i,m}(\bm{c} ^{i,m})$ according to equation(\ref{opt1}); \\
\quad \quad\quad ~If $m=M$, go to step 8. Otherwise, go to step 3.\\
\STATE \textit{Step 8:} Let $i = i+1$, and go to step 2 until $i= I_{max}$\\
\ENSURE ~~{$G^{i,M}(\bm{c} ^{i})$, $\bm{c}^{i,M}$};
\end{algorithmic}
\end{algorithm}

Based on this, we adopt  AGO algorithm to solve the original problem $\mathcal{P}1$. The details of the proposed iterative algorithm are given in Algorithm 2. In AGO algorithm, the D2D offloading gain increases after optimizing each $c_m$ as shown in step 7, and the convergence of AGO algorithm is guaranteed.
The computation in each iteration is dominated by step 5, where solution of transcendental equation $g_m(x) =0$ is achieved by root-finding algorithm. For simplicity, we adopt the bisection method \cite{bazaraa2013nonlinear} to find the root, and the maximum iteration number of bisection search is denoted by $K_b$. Therefore, the total computation complexity is $\mathcal{O}(I_{max}MK_b)$.

The convergence of AGO algorithm is proved as follows.
Let $c^i_m$ be the optimal pushing probability for group $\mathcal{G}_m$ in $i$ th iteration.
Then the offloading gain is obtained as $G^{i,m}(c^i_m|\bm{c}_{-m}^i)$ in step 7.
According to step 4 and step 6, $G^{i,m}(c^i_m|\bm{c}_{-m}^i)$ can be rewritten as
\begin{equation}
G^{i,m}(c^i_m|\bm{c}_{-m}^i)=G^{i,m}(c_1^{i},\cdots,c_m^i,c_{m+1}^{i-1},\cdots,c_{M}^{i-1})=G^{i,m+1}(c_{m+1}^{i-1}|\bm{c}_{-(m+1)}^i)
\end{equation}
After updating the pushing probability $c^i_{m+1}$ for group $\mathcal{G}_{m+1}$, the offloading gain is denoted by $G^{i,m+1}(c^i_{m+1}|\bm{c}_{-(m+1)}^i)$.
Therefore, we have the following inequality.
\begin{equation} \label{Astep}
G^{i,m+1}(c^i_{m+1}|\bm{c}_{-(m+1)}^i) \geq G^{i,m+1}(c_{m+1}^{i-1}|\bm{c}_{-(m+1)}^i)=G^{i,m}(c^i_m|\bm{c}_{-m}^i)
\end{equation}
The reason is that $c^i_{m+1}$ is obtained by proposition \ref{Theorem6}, which is the optimal pushing probability for group $\mathcal{G}_{m+1}$ given $\bm{c}_{-(m+1)}^i$ in $i$ th iteration.
However, the $c^{i-1}_{m+1}$ is obtained as the optimal solution for group $\mathcal{G}_{m+1}$ in $i-1$ th iteration, which is not necessarily optimal in current iteration.
Hence, in each iteration of AGO algorithm, the offloading gain is non-decreasing after each group updates its pushing probability at step 5.
Furthermore, the objective function of problem $\mathcal{P}1$ is upper-bounded by
\begin{equation}
G(\bm{c}) \leq \sum \limits_{m \in \mathcal{M}} t_m (1-c_m)  \leq  \sum \limits_{m \in \mathcal{M}} t_m
\end{equation}
Thus,  AGO algorithm is convergent.

\section{Simulation Results}

In this section, we carry out simulations to evaluate the offloading performance achieved by the optimal pushing strategy in the \textit{Group Independent Case} and the proposed AGO algorithm. We consider a single cell scenario where UEs density $\lambda_m$ of each group is set to be 0.05 UE per $m^2$. The D2D communication range is $r = 5m$.

\subsection{Offloading Performance in \textit{Group Independent Case}}
In this part, we study the offloading performance achieved by the optimal pushing strategy in the \textit{Group Independent Case}, and the impacts of sharing probability and content request probability of each group on the optimal pushing strategy.

\subsubsection{Offloading Performance}

\begin{figure}
\begin{minipage}[t]{0.5\textwidth}
\centering
\includegraphics[width=1\textwidth]{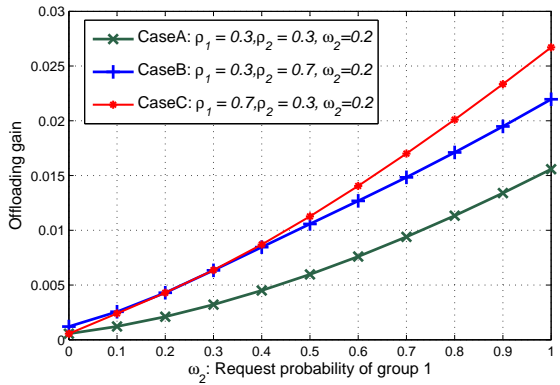}
\caption{Offloading gain performance versus $w_1$ with $M$=2}
\label{fig1}
\end{minipage}
\begin{minipage}[t]{0.5\textwidth}
\centering
\includegraphics[width=1\textwidth]{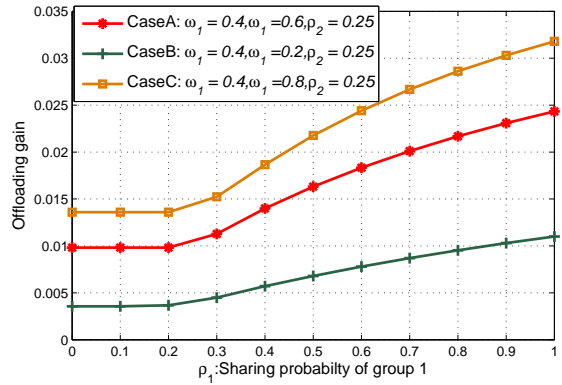}
\caption{Offloading gain performance versus $\rho_1$ with $M$=2}
\label{fig2}
\end{minipage}
\end{figure}

Fig. \ref{fig1} shows the system offloading gain $G$ versus the content request probability in group 1, where three cases with different system configurations are considered.
$\rho_1$ represents the sharing probability in group 1, and $\rho_2$ and $w_2$ represent the sharing probability and the request probability in group 2, respectively.
In general, when there is an increasing number of UEs requesting the content, the offloading gain grows in all considered cases, which benefits from the increased UE-Ts' density and D2D transmission success probability. Therefore, more UEs can get this content via D2D transmission.
Compared with other cases, Case A has the lowest offloading gain, which results from the small sharing probabilities. When $w_1 =0$, only group 2 has the interested UEs. At this point, Case B has a larger offloading gain compared with Case A and Case C, which is due to the higher sharing probability (0.7) of group 2. However, after $w_1$ reaches 0.3, the offloading gain in Case B is less than that of Case C. This is because most of UEs having interests are from group 1 in this interval. In Case B, most of the interested UEs are from a low sharing group, while the majority of requests are from a high sharing group in Case C. In Case B, more pushing efforts are made towards the low sharing group. Therefore, the offloading gain in Case C is better, since there are more content holders from a high sharing group.

Fig. \ref{fig2} shows the system offloading gain $G$ versus the sharing probability in group 1. In Fig. \ref{fig2}, when $\rho_1$ is less than $\rho_2$, the system offloading gain stays the same in all the considered cases. The reason is that the pushing effort is made to the UEs in group 2, and no pushing is made to the group 1 in this interval. The D2D communications are supplied by the content holders in group 2, since they are more willing to share. Only when the sharing probability of group 1 is larger than group 2, the UEs in group 1 will receive pushing from BS. This indicates that the emphasis of pushing is always placed in the group with higher sharing probability. After $\rho_1$ is larger than $\rho_2$, the system offloading gain increases with $\rho_1$. This is because most of the pushing effort is made to group 1 in this interval. Thus, the growth of $\rho_1$ makes the UE-Ts easier to find D2D helpers, so that the system offloading gain increases.
The offloading gain in Case B is less than that of Case A, which results from the smaller request probability in group 2. Similarly, since the request probability of group 2 in Case C is largest, the offloading gain in Case C is the largest. It is interesting to see that the offloading gain increases no matter whether the increased requests belong to a high sharing group. This implies that the proposed pushing strategy is attractive for distributing the ``popular" contents.

\subsubsection{Optimal Pushing Strategy}

\begin{figure}
\begin{minipage}[t]{0.5\textwidth}
\centering
\includegraphics[width=1\textwidth]{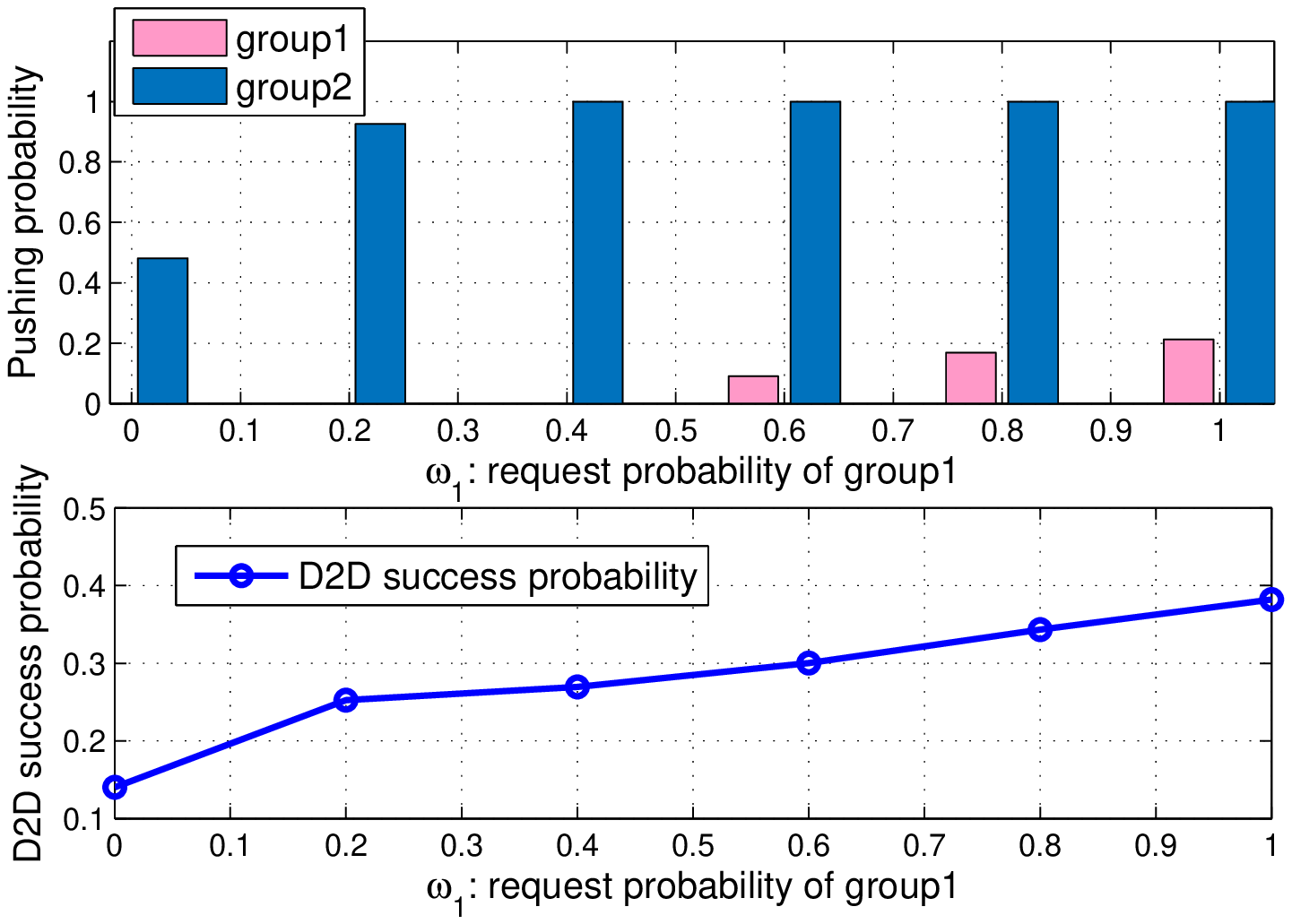}
\caption{Optimal pushing strategy versus $w_1$ with $\rho_1\!\!=\!\!0.2$,  $\rho_2 = 0.4,w_2 = 0.2$}
\label{fig3}
\end{minipage}
\begin{minipage}[t]{0.5\textwidth}
\centering
\includegraphics[width=1\textwidth]{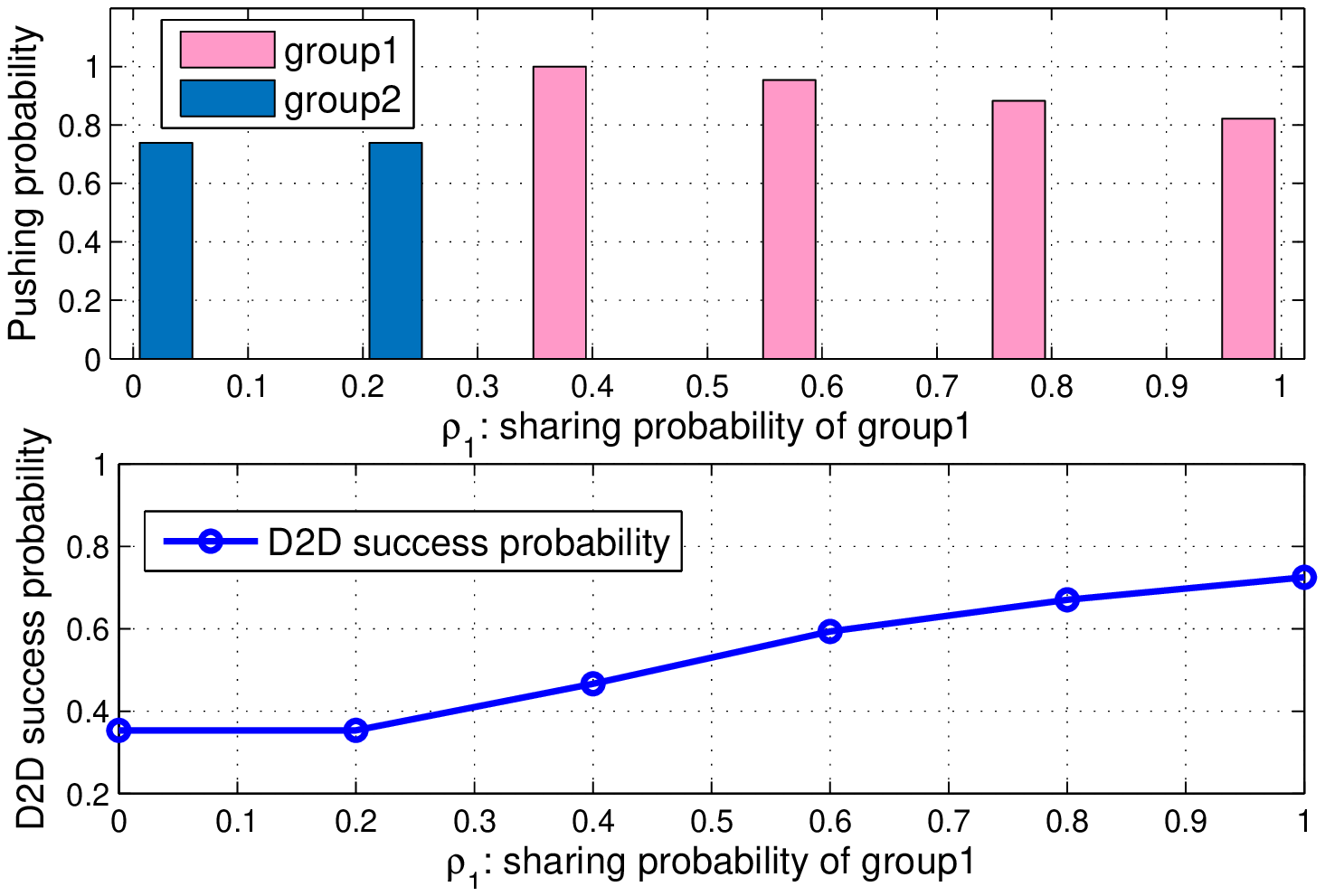}
\caption{Optimal pushing strategy versus $\rho_1$ with $w_1=0.4,w_2 = 0.6,\rho_2 = 0.25$}
\label{fig4}
\end{minipage}
\end{figure}

%

Fig. \ref{fig3} shows the optimal pushing probability in each group and the associate D2D success probability versus the content request density in group 1. In the \textit{Group Independent Case}, the D2D success probability is the same for each group. Thus, the subscript of $\mathbb{P}_m$ is ignored and it is rewritten as  $\mathbb{P}$. In Fig. \ref{fig3}, when $w_1$ increases, the pushing probability in group 2 increases to 1 due to the higher sharing probability. When $w_1$ is larger than 0.4, the number of UE-As in group 2 is not large enough to cope with the requests from increased number of UE-Ts in group1. This leads to the increase of pushing probability of group 1.  For the D2D success probability $\mathbb{P}$, it increases due to more pushing efforts are made. The increasing rate of $\mathbb{P}$ becomes slow when $w_1>0.2$, because most of the interested UEs in the high sharing group have already got the pushing, and the increased pushing efforts are made to UEs in low sharing group.

Fig. \ref{fig4} shows the optimal pushing strategy versus the sharing probability in group 1. When $\rho_1 < \rho_2$, the pushing effort is made in group 2 due to a higher sharing probability. Only when $\rho_1 > \rho_2$, the pushing effort is changed to group 1. However, it is interesting to see that the pushing probability decreases with the growth of $\rho_1$. This is because the content holders are more willing to offer D2D transmission, and the pushing effort can be saved. For the D2D success probability $\mathbb{P}$, it stays the same when $\rho_1 < \rho_2$ because the pushing probability is not changed. Since the sharing probability in group 1 increases, $\mathbb{P}$ keeps to go up, though the pushing probability decreases.

\vspace{-1.0em}
\subsection{AGO Algorithm Performance}
In this part, we will study the proposed AGO algorithm performance in terms of convergence property and the impact of initializations due to the nonconvexity of the original problem $\mathcal{P}1$. For each system configuration, content request probability $w_m$ is randomly drawn from uniform distribution in interval [0,1]. Given the constraint that $\rho_m^o \leq \rho_m^i$, the intergroup sharing probability $\rho_m^o$ and intragroup sharing probability $\rho_m^i$ are randomly drawn from uniform distribution in interval [0,0.3] and [0.7,1], respectively.

\subsubsection{Convergence Property}

\begin{figure}
\begin{minipage}[t]{0.5\textwidth}
\centering
\includegraphics[width=1\textwidth]{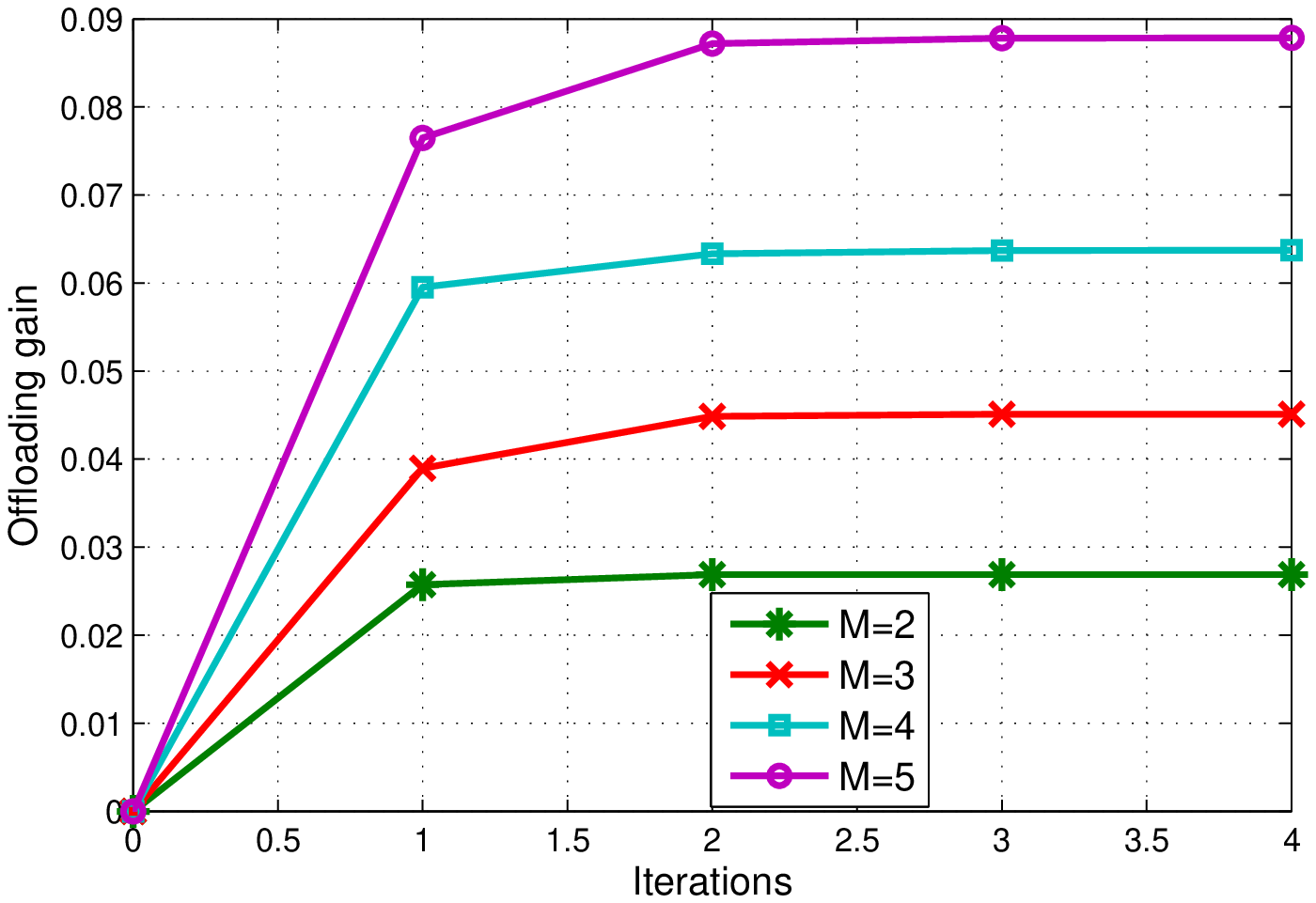}
\caption{Convergence property of AGO algorithm}
\label{fig5}
\end{minipage}
\begin{minipage}[t]{0.5\textwidth}
\centering
\includegraphics[width=1\textwidth]{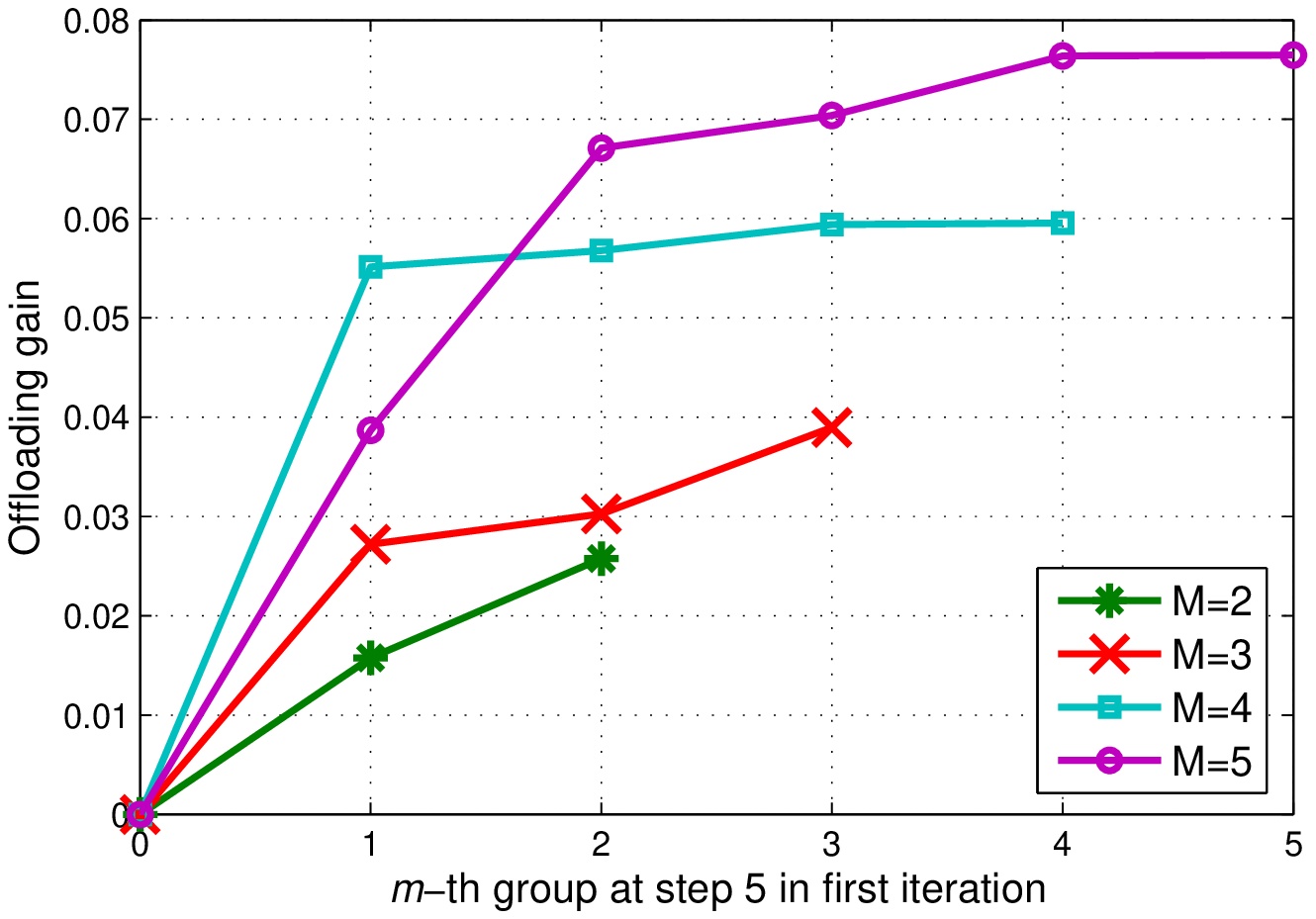}
\caption{Convergence behavior in the first iteration of AGO algorithm}
\label{fig6}
\end{minipage}
\end{figure}

Fig. \ref{fig5} shows the total offloading gain versus the number of iterations for AGO algorithm with different number of groups. As shown in Fig. \ref{fig5}, the total offloading gain increases with iterations. The algorithm converges rapidly to the optimum value within 2 iterations for all considered cases, which makes this method attractive for implementation. Therefore, we set the maximum iteration number to be 2 in the following simulations.

Fig. \ref{fig6} illustrates the convergence behavior at the first iteration of  AGO algorithm.
As expected, we can see that the offloading gain increases after each group updates its pushing probability in step 5.
This result also verifies the theoretical results given in proposition \ref{Theorem6}.

\subsubsection{Impact of Initialization}

\begin{figure}
\begin{minipage}[t]{0.5\textwidth}
\centering
\includegraphics[width=1\textwidth]{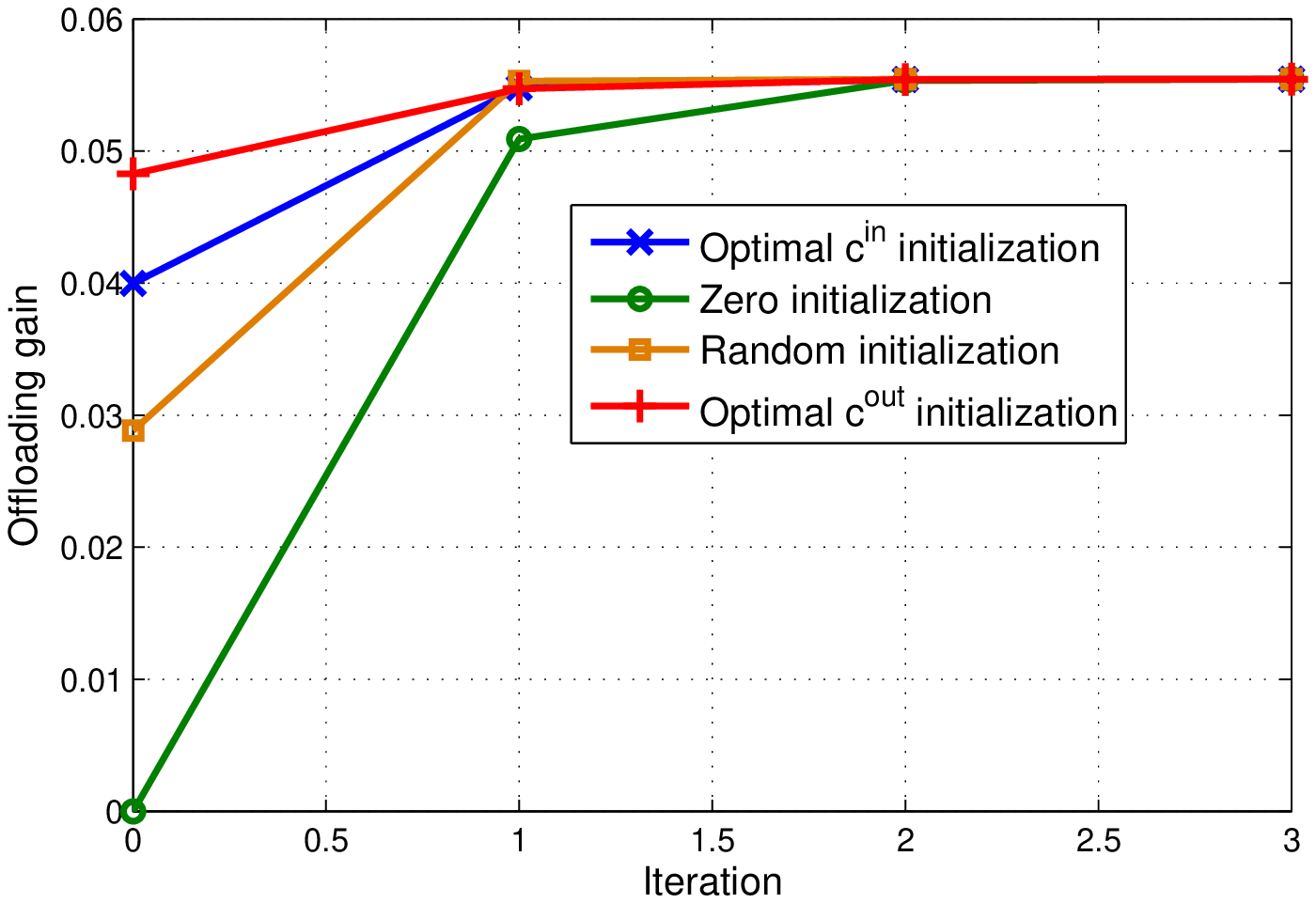}
\caption{Impact of initialization on AGO algorithm}
\label{fig7}
\end{minipage}
\begin{minipage}[t]{0.5\textwidth}
\centering
\includegraphics[width=1\textwidth]{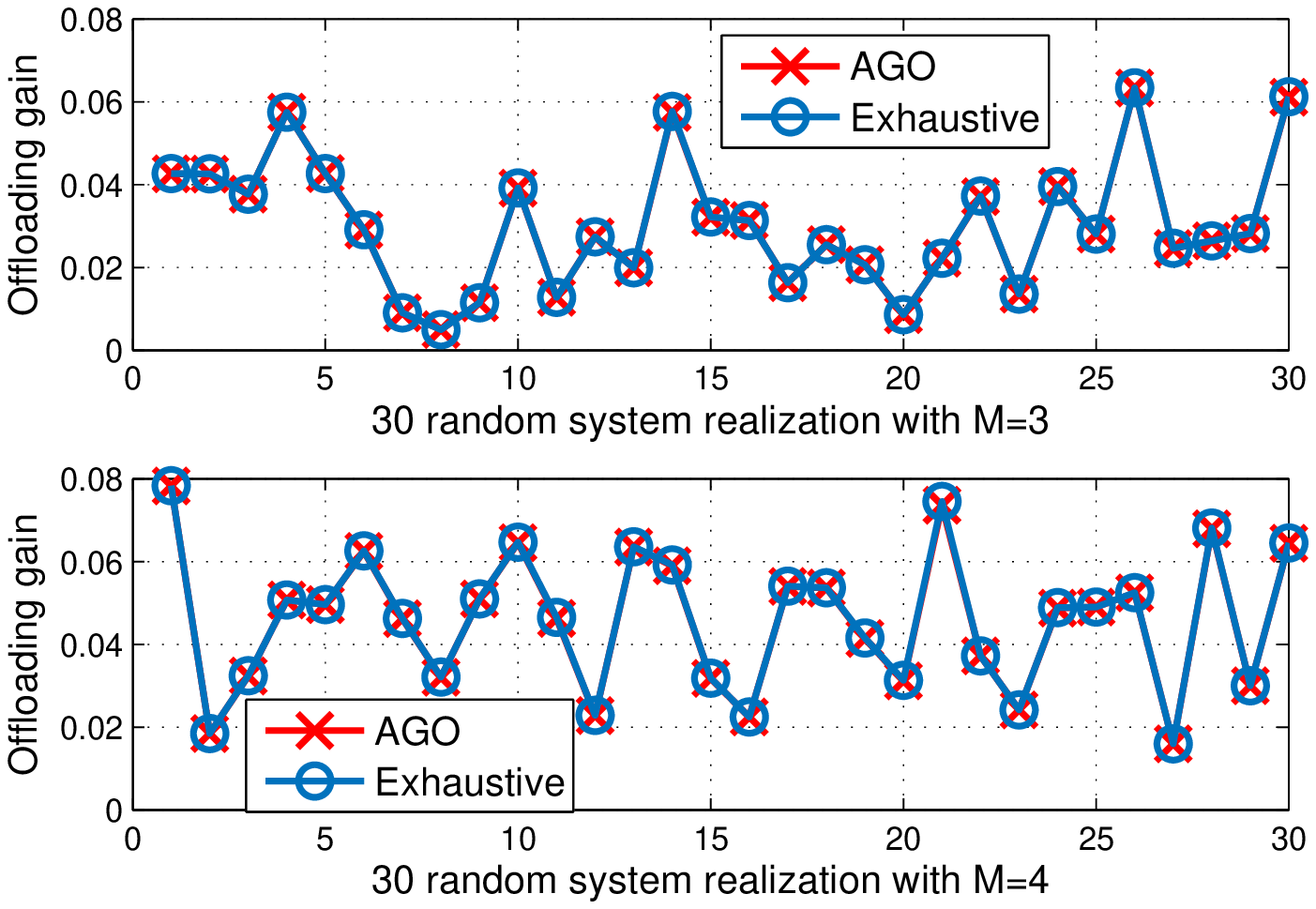}
\caption{Performance comparison of AGO algorithm and exhaustive search}
\label{fig8}
\end{minipage}
\end{figure}

Although the convergence of  AGO algorithm is guaranteed, different initializations may result in different local maximum due to the nonconvexity of problem $\mathcal{P}1$. To test the impact of different initializations, four methods are investigated for $M=3$.
In Fig. \ref{fig7}, $\bm{c}^{out}$ represents the optimal pushing strategy achieved in \textit{Group Independent Case} where only intergroup sharing probabilities are considered, i.e., $\rho_m=\rho_m^o$. Similarly, $\bm{c}^{in}$ represents the optimal pushing strategy achieved in \textit{Group Independent Case} where $\rho_m=\rho_m^i$.
In zero initialization, the pushing probability of each group is set to be zero.
In random initialization, the pushing probability of each group is randomly generated. In Fig. \ref{fig7}, AGO algorithm converges to the same optimum for all considered cases within 2 iterations, which shows that  AGO algorithm is not sensitive to the initialization value. In addition, Fig. \ref{fig7} shows that $\bm{c}^{out}$ initialization is closer to the potential optimum compared with other methods.

Fig. \ref{fig8} compares the performance of  AGO algorithm and exhaustive search method in 30 randomly generated system realizations. The pushing strategy in AGO algorithm is initialized by $\bm{c}^{out}$, and the iteration number is set to be 2. The search step size for pushing probability of each group in exhaustive search is $0.001$, and its total complexity is $(1001)^M$.
It can be seen that the converged optimum is almost the same as that of exhaustive search, which can be approximately viewed as the globally optimum.

\section{Conclusion}
In this paper, with the consideration of content preference and sharing willingness of the human users, we investigated the optimal pushing strategy to maximize the system offloading gain. UEs are classified into groups according to their content preferences, and shared content with intergroup and intragroup UEs at different sharing probabilities.
In the content dissemination process, only the UEs having interests in the content will accept the content pushing. The D2D transmissions are affected by the different sharing probabilities between groups.
By optimizing the pushing probability in each group, the system offloading gain in terms of offloaded traffic was maximized. Although the optimization problem is nonconvex, the closed-form optimal pushing strategy was achieved in a special case when intergroup sharing probability is same with the intragroup sharing probability. In addition, an alternative group optimization algorithm, AGO, was proposed to solve the general case of the optimization problem. Finally, the simulation results showed that the converged result of AGO algorithm is near the global optimum. The impacts of the content preference and sharing willingness on the optimal pushing strategy were also uncovered. Specifically, the following insights can be observed.
\begin{itemize}
  \item The sharing willingness of human users is crucial to offloading performance. Therefore, the pushing effort should be focused on people who are more willing to share for them to carry out D2D assisted content offloading.
\item When the majority of the interested people have low sharing willingness, the offloading gain achieved is less as compared to the opposite case, where majority of interested users are willing to share the content, even though more pushing effort is made.
\end{itemize}

When the proposed offloading algorithm is applied to the multiple contents scenario, a coupling effect will be considered due to the limited energy budget of UEs. Moreover,  the frequency resource allocation and interference management could be integrated into the proposed scheme, which will be investigated in our future work.

\appendices
\section{Proof of Lemma \ref{Theorem1}}
\setcounter{equation}{0}
\renewcommand{\theequation}{\thesection.\arabic{equation}}
Assuming that there exist two different groups $\mathcal{G}_i$ and $\mathcal{G}_j$, such that their optimal pushing probabilities $c_i^{\ast}$  and $c_j^{\ast}$ are between zero to one, i.e., $0 < c_i^{\ast} < 1$ and $0 < c_j ^{\ast}< 1$.
Since the objective function $G_{f}(\bm{c})$ of problem $\mathcal{P}2$ has the maximum value at point $\bm{c}^{\ast}$,  $\bm{c}^{\ast}$ is the solution of the following equation set:
\begin{equation}
\frac{\partial G_{f}(\bm{c}) }{\partial c_i} = 0,
\frac{\partial G_{f}(\bm{c}) }{\partial c_j} = 0.
\end{equation}

To make it more clear, the equation set is rewritten as
\begin{eqnarray}
&1+B\rho_i \sum \limits_{k \in \mathcal{M}} t_k (1-c_k) = \textrm{exp}\left({B\sum \limits_{k \in \mathcal{M}}t_k \rho_k c_k }\right), \\
&1+B\rho_j \sum \limits_{k \in \mathcal{M}} t_k (1-c_k) = \textrm{exp}\left({B\sum \limits_{k \in \mathcal{M}}t_k \rho_k c_k }\right).
\end{eqnarray}
In the \textit{non-uniform} sharing scenario, we have $\rho_i \neq \rho_j $, so that the equation set has no feasible solution. Hence, the assumption is invalid, and there is at most one group $\mathcal{G}_i$ in which the optimal $c_i^{\ast}$ is between $0$ and $1$.
$\hfill\blacksquare$

\section{Proof of Lemma \ref{Theorem2}}
\setcounter{equation}{0}
\renewcommand{\theequation}{\thesection.\arabic{equation}}
For problem $\mathcal{P}2$, it is not hard to verify that the linear independence constraint qualification (LICQ) \cite{eustaquio2008constraint} holds for all feasible pushing strategies. Therefore, the qualification applies at the global optimum. This implies that the Karush-Kuhn-Tucker (K.K.T) conditions are necessary conditions for the global optimum.
The Lagrangian associated with problem $\mathcal{P}2$ is written as
\begin{equation}
\mathcal{L}(\bm{\alpha},\bm{\beta},\bm{c})
=\sum \limits_{m \in \mathcal{M}} t_m  (1- c_m)
\left ( 1 - \textrm{exp}{(-B\sum\limits_{k \in \mathcal{M}} t_k \rho_k c_k )} \right)
+ \sum \limits_{m \in \mathcal{M}}\alpha_m c_m
- \sum \limits_{m \in \mathcal{M}}\beta_m (c_m-1).
\end{equation}
where $\bm{\alpha}=[\alpha_1, \alpha_2, \cdots, \alpha_M]$, $\bm{\beta} =[\beta_1, \beta_2, \cdots, \beta_M]$. $\alpha_m$ and $\beta_m$ are the non-negative dual variables associated with the constraints $c_m \geq 0$ and $c_m-1 \leq 0$, respectively.

Thus, we have the following K.K.T conditions.
\begin{eqnarray}
\frac{\partial \mathcal{L}(\bm{\alpha},\bm{\beta},\bm{c})}{\partial c_m} &=& 0,
\forall m \in \mathcal{M} ,  \label {Km1}\\
\alpha_m c_m &=&0 ,\forall m \in \mathcal{M},\\
\beta_m (c_m-1) &=&0,\forall m \in \mathcal{M},\\
\alpha_m \geq 0 ,\beta_m &\geq &0 ,\forall m \in \mathcal{M},\\
0 \leq c_m &\leq& 1,\forall m \in \mathcal{M}.
\end{eqnarray}

From (\ref {Km1}), we can obtained that
\begin{equation} \label {LDiv}
\textrm{exp}{(-B\sum\limits_{k \in \mathcal{M}} t_k \rho_k c_k )}\left( B \rho_m\sum \limits_{k \in \mathcal{M}} t_k  (1- c_k) + 1\right) - 1=\frac{1}{t_m} (\beta_m-\alpha_m) , \textrm{for all} m \in \mathcal{M}.
\end{equation}

At the global optimal pushing strategy, if $0<c_m^{\ast} <1$, then $\alpha_m=0$, and $\beta_m=0$.
Therefore, following equation holds,
\begin{equation} \label {K1Op}
 B \rho_m\sum \limits_{k \in \mathcal{M}} t_k  (1- c_k) + 1 = \textrm{exp}{(B\sum\limits_{k \in \mathcal{M}} t_k \rho_k c_k )}.
\end{equation}

For the group $\mathcal{G}_{i}$, from its associated K.K.T condition $\frac{\partial \mathcal{L}}{\partial c_{i}} =0$, the following equation holds in the global optimal pushing strategy,
\begin{equation} \label{K2Op}
\textrm{exp}{(-B\sum\limits_{k \in \mathcal{M}} t_k \rho_k c_k )}\left( B \rho_i\sum \limits_{k \in \mathcal{M}} t_k  (1- c_k) + 1\right) - 1=\frac{1}{t_i} (\beta_i-\alpha_i).
\end{equation}

If $\rho_i <\rho_m$, combing (\ref {K2Op}) with (\ref {K1Op}), it can be derived that
\begin{equation}
\beta_i-\alpha_i <0.
\end{equation}
Given the fact that the dual variables are non-negative, thus $\alpha_i >0$ and $c_i^{\ast} =0$.
Similarly, for group $\mathcal{G}_{j}$, if $\rho_{j} >\rho_{m}$, from its associated K.K.T condition that $\frac{\partial \mathcal{L}}{\partial c_j} =0$, we can obtain that
\begin{equation}
\beta_j-\alpha_j >0.
\end{equation}
Therefore, it can be inferred that $\beta_j>0$ and $c_j^{\ast} =1$.
$\hfill\blacksquare$

\section{Proof of Corollary \ref{Corcm}}
\setcounter{equation}{0}
\renewcommand{\theequation}{\thesection.\arabic{equation}}
According to Theorem \ref{Theorem2}, for each group $\mathcal{G}_{i}$ with $1\leq i \leq m-1$, the optimal pushing probability is $c_{i}^{\ast} =0$. For each group $\mathcal{G}_{j}$ with $m+1 \leq j \leq M$, the optimal pushing probability is $c_{j}^{\ast} =0$, so that (\ref{K1Op}) is simplified to
\begin{equation} \label{EquCm}
B \rho_m({\sum\limits_{i=1}^{m}t_i}-t_m c_m^{\ast}) + 1= \textrm{exp}\left({\sum\limits_{j=m+1}^{M} B t_j \rho_j + B t_m\rho_m c_m^{\ast}}\right).
\end{equation}
By employing the following equation that
\begin{equation}
e^{ax+b} = cx+d \to x = -\frac{d}{c}-\frac{1}{a}\mathcal{W}(-\frac{a}{c} e^{b-\frac{ad}{c}}),
\end{equation}
where $\mathcal{W}$ is the Lambert-W function, the equation (\ref{cm}) is obtained.
$\hfill\blacksquare$

\section{Proof of Theorem \ref{Theorem3}}
\setcounter{equation}{0}
\renewcommand{\theequation}{\thesection.\arabic{equation}}

\begin{lemma}\label{proUni}
Assuming that $M$ groups are sorted in the order $\rho_1 < \rho_2 <\cdots < \rho_M$,
then there is at most one group that satisfies the conditions  (\ref{1Condition}) and (\ref{0Condition}) at the same time.
\end{lemma}
Lemma \ref{proUni} is introduced to facilitate the proof, where its proof is provided in Appendix E.

The proof of Theorem \ref{Theorem3} goes by contradiction for both the ``\textit{if}" part and the ``\textit{only if}" part.

First consider the proof of the ``\textit{if}" part.
If the conditions in (\ref{1Condition}) and (\ref{0Condition}) hold for group $\mathcal{G}_m$ at the same time, we suppose that group $\mathcal{G}_m$ is not the ``\textit{watershed}"  group. According to Lemma \ref{Theorem1}, the optimal pushing probability of group $\mathcal{G}_m$ is either $c_m^{\ast}=1$ or $c_m^{\ast}=0$. In the following part, we will show that the assumption $c_m^{\ast}=1$ contradicts condition (\ref{0Condition}) and $c_m^{\ast}=0$ contradicts condition (\ref{1Condition}).

If $c_m^{\ast}=1$, according to Corollary \ref{Cor2}, it is inferred that $c_j^{\ast}=1$ for all groups $\mathcal{G}_j$ with $m+1\leq j \leq M$. Therefore, the optimal pushing strategy is
$\bm{c}^{\ast} =[c_1^{\ast}, \cdots, c_{m-1}^{\ast},\underbrace{1, \cdots, 1}_{M-m+1}]$, and the K.K.T condition (\ref {LDiv}) is reduced to
\begin{equation}\label{contr1}
\textrm{exp}\left(-B \sum\limits_{i= 1}^{m-1} t_i \rho_i c_i^{\ast}-B \sum\limits_{j= m}^{M} t_j \rho_j\right)\left( B \rho_m\sum\limits_{i= 1}^{m-1} t_i (1-c_i^{\ast}) + 1\right) - 1=\frac{1}{t_m} (\beta_m-\alpha_m).
\end{equation}
In addition, since $c_m^{\ast}=1$, then $\alpha_m=0$ and $\beta_m\geq 0$. The following inequality is inferred from (\ref{contr1}).
\begin{equation}\label{asp1}
 B \rho_m\sum\limits_{i= 1}^{m-1} t_i (1-c_i^{\ast}) + 1 \geq \textrm{exp}\left(B \sum\limits_{i= 1}^{m-1} t_i \rho_i c_i^{\ast}+B \sum\limits_{j= m}^{M} t_j \rho_j\right).
\end{equation}
Due to the fact that $0 \leq c_i \leq 1$, from the RHS of (\ref{asp1}), we obtain that
\begin{equation}\label{asp2}
\textrm{exp}\left(B \sum\limits_{i= 1}^{m-1} t_i \rho_i c_i^{\ast}+B \sum\limits_{j= m}^{M} t_j \rho_j\right)\geq \textrm{exp}(B \sum\limits_{j= m}^{M} t_j \rho_j).
\end{equation}
From the LHS of  (\ref{asp1}), the following inequality is obtained.
\begin{equation}\label{asp3}
 B \rho_m\sum\limits_{i= 1}^{m-1} t_i + 1 \geq B \rho_m\sum\limits_{i= 1}^{m-1} t_i (1-c_i^{\ast}) + 1.
\end{equation}

Substituting (\ref{asp3}) and (\ref{asp2}) in (\ref{asp1}), the following inequality is obtained, which contradicts condition (\ref{0Condition}).
\begin{equation}
 B \rho_m\sum\limits_{i= 1}^{m-1} t_i + 1 \geq \textrm{exp}(B \sum\limits_{j= m}^{M} t_j \rho_j).
\end{equation}

If $c_m^{\ast}=0$, according to Corollary \ref{Cor2}, we infer that $c_i^{\ast}=0$ for all groups $\mathcal{G}_i$ with $1\leq i \leq m-1$. Therefore, the optimal pushing strategy is $\bm{c}^{\ast} =[\underbrace{0, \cdots, 0}_{m}, c_{m+1}^{\ast}, \cdots, c_{M}^{\ast}]$. The K.K.T condition in (\ref {LDiv}) is reduced to
\begin{equation}\label{contr2}
\textrm{exp}(-B \sum\limits_{j=m+1}^{M} t_j \rho_j c_j^{\ast})\left( B \rho_m\sum\limits_{j=m+1}^{M} t_j (1-c_j^{\ast}) +B \rho_m\sum\limits_{i=1}^{m} t_i + 1\right) - 1=\frac{1}{t_m} (\beta_m-\alpha_m).
\end{equation}
In addition, since $c_m^{\ast}=0$, then $\alpha_m\geq 0$ and $\beta_m=0$, the following inequality is inferred from (\ref{contr2}).
\begin{equation}\label{asump1}
B \rho_m\sum\limits_{j= m+1}^{M} t_j (1-c_j^{\ast}) +B \rho_m\sum\limits_{i=1}^{m} t_i + 1 \leq \textrm{exp}\left(B \sum\limits_{j= m}^{M} t_j \rho_j c_j^{\ast}\right).
\end{equation}
Due to the fact that $0 \leq c_j \leq 1$, the following inequalities are obtained from the RHS and the LHS of (\ref{asump1}), respectively.
\begin{equation}
\textrm{exp}\left(B \sum\limits_{j= m}^{M} t_j \rho_j c_j^{\ast}\right) \leq \textrm{exp}\left(B \sum\limits_{j= m}^{M} t_j \rho_j\right), 
B \rho_m\sum\limits_{j= m+1}^{M} t_j (1-c_j^{\ast}) +B \rho_m\sum\limits_{i=1}^{m} t_i + 1 \geq B \rho_m\sum\limits_{i=1}^{m} t_i + 1 \label{asump3}.
\end{equation}
Combine (\ref{asump3}) with  (\ref{asump1}), the following inequality is readily obtained.
\begin{equation}\label{contrd2}
B \rho_m\sum\limits_{i=1}^{m} t_i + 1 \leq  \textrm{exp}\left(B \sum\limits_{j= m}^{M} t_j \rho_j\right).
\end{equation}
Obviously, (\ref{contrd2}) contradicts the condition (\ref{1Condition}).

Overall, if the conditions in (\ref{1Condition}) and (\ref{0Condition}) hold for group $\mathcal{G}_m$ at the same time, the optimal solution of group $\mathcal{G}_m$ is larger than 0 and less than 1. Therefore, the optimal solution of $\mathcal{P}2$ is obtained as $\bm{c}^{\ast} =[\underbrace{0, \cdots, 0}_{m-1},c_m^{\ast},\underbrace{1, \cdots, 1}_{M-m}]$, where $c_m^{\ast}$ is given by (\ref{cm}).

Next, consider the ``\textit{only if}" part.
Suppose that $\bm{c}^{\ast}=[\underbrace{0, \cdots, 0}_{m-1},c_m^{\ast},\underbrace{1, \cdots, 1}_{M-m}]$ is the optimal pushing strategy, meanwhile condition (\ref{1Condition}) or condition (\ref{0Condition}) is not satisfied. Nevertheless, the following proof proves that there exists a different pushing strategy which achieves a larger offloading gain than $\bm{c}^{\ast}$.
The offloading gain achieved by $\bm{c}^{\ast}$ is
\begin{equation}\label{Escm}
G_{f}(\bm{c}^{\ast})=\left(\sum\limits_{i=1}^{m} t_i - t_m c_m^{\ast}\right)\left(1-\textrm{exp}\left(-B\sum\limits_{j=m+1}^{M} t_j\rho_j-B t_m\rho_m c_m^{\ast}\right)\right).
\end{equation}

According to the proof of Corollary \ref{Corcm}, the following equation holds for $c_m^{\ast}$.
\begin{equation} \label{EquCmKKT}
B \rho_m\left({\sum\limits_{i=1}^{m}t_i}-t_m c_m^{\ast}\right) + 1= \textrm{exp}\left({\sum\limits_{j=m+1}^{M} B t_j \rho_j + B t_m\rho_m c_m^{\ast}}\right).
\end{equation}

Substituting the RHS and LHS of equation (\ref{EquCmKKT}) into (\ref{Escm}), we have the following two equivalent expressions, respectively.
\begin{eqnarray}
G_{f}(\bm{c}^{\ast})=\frac{1}{B\rho_m}(\textrm{exp}(R_m)+\textrm{exp}(-R_m)-2),
G_{f}(\bm{c}^{\ast})= \frac{B\rho_m L_m^2}{B\rho_m L_m +1}.
\end{eqnarray}
where $L_m =\sum\limits_{i=1}^{m} t_i - t_m c_m^{\ast} $ and $R_m = \sum\limits_{j=m+1}^{M} B t_j\rho_j+B t_m\rho_m c_m^{\ast}$.

If the condition (\ref{1Condition}) is not satisfied, the following inequality holds.
\begin{equation}
1+B \rho_m \sum\limits_{i=1}^{m} t_i \leq \textrm{exp}\left(B\sum\limits_{j=1+m}^{M} t_j\rho_j\right). \label {Only0}\\
\end{equation}
We define a feasible pushing strategy as $\bm{c}^0 =[\underbrace{0, \cdots, 0}_{m},\underbrace{1, \cdots, 1}_{M-m}]$. It is easy to verify that the objective value of $\mathcal{P}2$ achieved by $\bm{c}^0$ is
\begin{equation}
G_{f}(\bm{c}^{0}) = \sum\limits_{i=1}^{m} t_i\left(1-\textrm{exp}\left(-B\sum\limits_{j=m+1}^{M} t_j\rho_j\right)\right).
\end{equation}
Then, according to equation (\ref{Only0}), it can be inferred that
\begin{equation}\label{end}
G_{f}(\bm{c}^0) \geq \frac{B\rho_m \left(\sum\limits_{i=1}^{m} t_i\right)^2}{B\rho_m \sum\limits_{i=1}^{m} t_i +1}.
\end{equation}

Denote function $u(x)=\frac{a x^2}{a x +1}$. We have $u'(x)=\frac{a x^2+2ax}{(a x +1)^2}>0$, $\forall x >0$. Therefore, function $u(x)$ is a strictly increasing function. Since $\sum\limits_{i=1}^{m} t_i > L_m$, it can be obtained that
$G_{f}(\bm{c}^0) > G_{f}(\bm{c}^{\ast})$. Therefore, this contradicts with the presumption that $\bm{c}^{\ast}$ is the optimal pushing strategy.


If the condition (\ref{0Condition}) is not satisfied, we can find another feasible solution denoted by $\bm{c}^1 =[\underbrace{0, \cdots, 0}_{m-1},\underbrace{1, \cdots, 1}_{M-m+1}]$.
It is easy to verify that $G_{f}(\bm{c}^1) > G_{f}(\bm{c}^{\ast})$, which also contradicts that $\bm{c}^{\ast}$ is global optimum.
The proof is similar with the procedure from (\ref{Only0}) to (\ref{end}), and it is omitted for brevity.
Therefore, the $\bm{c}^{\ast} =[\underbrace{0, \cdots, 0}_{m-1},c_m^{\ast},\underbrace{1, \cdots, 1}_{M-m}]$ is the optimal solution of problem $\mathcal{P}2$, where $c_m^{\ast}$ is given by (\ref{cm}), \textit{only if} the two conditions (\ref{1Condition}) and (\ref{0Condition}) hold at the same time.

By combining the proofs of the ``\textit{if}" part and the ``\textit{only if}" part, Theorem \ref{Theorem3} is proved.
$\hfill\blacksquare$

\section{Proof of Lemma \ref{proUni}}
\setcounter{equation}{0}
\renewcommand{\theequation}{\thesection.\arabic{equation}}
We prove the uniqueness by contradiction. It is assumed that two different groups $\mathcal{G}_m$ and $\mathcal{G}_k$ both satisfy the two conditions (\ref {0Condition}) and (\ref {1Condition}).
Without loss of generality, it is assumed that $\rho_k > \rho_m $. In this case, since $M$ groups are sorted in the order $\rho_1 < \rho_2 <\cdots < \rho_M$, then $k-1 \geq m$. For group $\mathcal{G}_k$, the condition (\ref {0Condition}) is written as
\begin{eqnarray} \label{uq1}
 1+B \rho_k \sum\limits_{i=1}^{k-1} t_i < \textrm{exp}\left(B\sum\limits_{j=k}^{M} t_j\rho_j\right).
\end{eqnarray}
However, for the left hand side (LHS) of (\ref{uq1}), we have the following inequality
\begin{equation}
1+B \rho_m \sum\limits_{i=1}^{m} t_i <  1+B \rho_k \sum\limits_{i=1}^{k-1} t_i. \label{uq2}
\end{equation}
For the right hand side (RHS) of (\ref{uq1}), we have
\begin{equation}
\textrm{exp}\left(B\sum\limits_{j=k}^{M} t_j\rho_j\right)\leq\textrm{exp}\left(B\sum\limits_{j=m+1}^{M} t_j\rho_j\right). \label{uq3}
\end{equation}
Substituting (\ref{uq2}) and (\ref{uq3}) into (\ref{uq1}), the following inequality is obtained.
\begin{eqnarray}\label{ctr1}
1+B \rho_m \sum\limits_{i=1}^{m} t_i <\textrm{exp}(B\sum\limits_{j=m+1}^{M} t_j\rho_j).
\end{eqnarray}
However, according to the above assumption, condition (\ref {1Condition}) also holds for group $\mathcal{G}_m$, which is contradictory to (\ref {ctr1}).
When $\rho_k < \rho_m $, similar proof will follow.
Therefore, when the sharing probabilities are different by different groups, there is at most one group that satisfies (\ref{1Condition}) and (\ref{0Condition}) at the same time.
$\hfill\blacksquare$

\vspace{-2em}
\section{Proof of Theorem \ref{Theorem5}}
\setcounter{equation}{0}
\renewcommand{\theequation}{\thesection.\arabic{equation}}
Sort the $M$ groups in the  ascending order of $\rho_m$, i.e.,  $\rho_1 <\cdots <\rho_{k_1}= \cdots =\rho_{k_n}<\cdots< \rho_M$. Let $\mathcal{K}=\{k_1,\cdots,k_n\}$ denotes the set of groups with the same sharing probability. We define a new group 0 with sharing probability that $\rho_0 =\rho_{k_1}=\cdots =\rho_{k_n}$, and the request density of this group 0 is denoted as $t_0$, which is given by
\begin{equation} \label{group0}
t_0 = \sum\limits_{k=k_1}^{k_n} t_k,c_0 =\frac{1}{t_0}\sum\limits_{k=k_1}^{k_n} t_kc_k.
\end{equation}
where $c_0$ is the pushing probability of group 0.

By substituting group 0 for the groups in $\mathcal{K}$
, problem $\mathcal{P}2$ is reduced to
\begin{subequations}
\begin{align}
 \mathcal{P}2.1:\mathop{\max }_{c_m,c_0}
                  &\quad   G_{f} = \left(\sum \limits_{m \in \mathcal{M}'} t_m  (1- c_m)+t_0(1-c_0) \right)
                                  \left ( 1 - \textrm{e}^{-B\sum \limits_{m \in \mathcal{M}'}t_m \rho_m c_m-Bt_0 \rho_0 c_0}  \right),       \label{optU}   \\
\textrm{s.t.} &\quad 0\leq c_m \leq 1 ,\quad m \in  \mathcal{M}', \\
&\quad 0\leq c_0 \leq 1.
\end{align}
\end{subequations}
where $\mathcal{M}'$ is set of the left $M-n$ groups, i.e., $\mathcal{M}' = \mathcal{M}-\mathcal{K}$. Obviously, problem $\mathcal{P}2.1$ is the same as the \textit{non-uniform} sharing scenario, except that total group number is changed to $M-n+1$, and the optimal solution can be obtained directly from Theorem \ref{Theorem4}.

We denote the optimal pushing strategy of group 0 as $c_0^{\ast}$. It is easy to verify that for a given $c_0^{\ast}$, there exists multiple $(c_{k_1}^{\ast},\cdots,c_{k_n}^{\ast})$ that satisfy the following condition.
\begin{equation}
t_0c_0^{\ast} =\sum\limits_{k=k_1}^{k_n} t_k c_k^{\ast}.
\end{equation}
For example, a special case is $c_{k_1}^{\ast}=\cdots=c_{k_n}^{\ast}=c_0^{\ast}$.
$\hfill\blacksquare$

\section{Proof of Proposition \ref{Theorem6}}
\setcounter{equation}{0}
\renewcommand{\theequation}{\thesection.\arabic{equation}}
Since $\mathcal{P}_3$ is a convex problem and satisfies the Slater's condition, we can solve $\mathcal{P}_3$ by solving its dual problem due to the zero gap between them \cite{boyd2004convex}. The Lagrangian associated with this problem is written as
\begin{equation}
\mathcal{L}^{m}(c_m, \gamma_m, \eta_m)=G_{-m}(c_m)  +\gamma_m c_m - \eta_m(c_m -1).
\end{equation}
where $\gamma_m$ and $\eta_m$ denote the dual variables associated with constraint $c_m \geq 0$ and $c_m \leq 1$, respectively.  The dual problem of $\mathcal{P}_3$ is given by
\begin{equation}
\mathop{\min }_{\gamma_m,\eta_m} \mathop{\sup }_{c_m} \mathcal{L}^{m}(c_m, \gamma_m, \eta_m).
\end{equation}
Accordingly, the K.K.T conditions are given by
\begin{eqnarray}
\frac{\partial \mathcal{L}^{m}(c_m, \gamma_m, \eta_m)}{\partial c_m} &=&0 \label {lm},\\
\gamma_m c_m &=&0,   \label{gama} \\
\eta_m (c_m-1) &=&0,     \label{eta}   \\
\gamma_m \geq 0 ,\eta_m &\geq &0, \\
c_m \geq 0,1-c_m &\geq& 0.
\end{eqnarray}

For simplicity, we define function
\begin{equation}
g_m(x) =(1+ B t_m \rho_m(1-x)) \textrm{exp}({-B t_m \rho_m x - \varphi_m}) + B I_m \rho_m^o \textrm{exp}(-{B t_m \rho_m^o x}) -1.
\end{equation}
where $I_m=\sum_{k \neq m}^M Q_k \textrm{exp}(- \Phi_k)$.

It is easy to verify that $g'_m(x) <0$, so that $g_m(x)$ is a monotonically decreasing function w.r.t $x$ for all $m \in \mathcal{M}$.
First, it follows that $\lim\limits_{x \to -\infty}g_m(x) \to +\infty >0$, and $\lim\limits_{x \to +\infty}g_m(x) = -1 <0$. This implies that function $g_m(x)=0$ has a unique solution for each $m \in \mathcal{M}$. The unique solution for a given $m$ is denoted by $\Gamma_m$, which can be calculated by the root-finding algorithms, e.g., the bisection method.

By taking the deritivative of the Lagrangian w.r.t $c_m$, (\ref {lm}) is given by
\begin{equation}
g_m(c_m) + \frac{1}{t_m}(\gamma_m - \eta_m) =0 \label{lm0}.
\end{equation}
Then we discuss the optimal solution satisfying K.K.T conditions in the three possible regions of $\Gamma_m$, i.e., $(-\infty,0)$, $[0,1]$, and $(1,+\infty)$, respectively.

\begin{itemize}
\item When $\Gamma_m \in (-\infty,0)$, we have $g_m(c_m) < 0$ for any $c_m \in [0,1]$.  Therefore, from (\ref{lm0}), we can infer that  $\gamma_m - \eta_m >0$. Since dual variables are non-negative, we can claim that $\gamma_m >0$ and the optimal pushing probability $c_m^{\ast}=0$.

\item When $\Gamma_m \in [0,1]$, due to the complementary slackness conditions in (\ref{gama}) and (\ref{eta}), it follows that $\gamma_m=\eta_m =0$. Thus, the optimal pushing probability $c_m^{\ast}= \Gamma_m$.

\item When $\Gamma_m \in (1,+\infty)$, $g_m(c_m) > 0$ for $c_m \in [0,1]$. Therefore, we have $\eta_m >0$ and $c_m^{\ast}=1$.
\end{itemize}

Overall, we can conclude that the optimal pushing probability for group $\mathcal{G}_m$ in problem $\mathcal{P}3$ is $[\Gamma_m]^1_0$, where $a=[x]^1_0$ means that if $x>1$, $a=1$; if $x<0$, $a=0$, and if $0\leq x \leq 1$, $a=x$. Thus, the Proposition \ref{Theorem6} is proved.
$\hfill\blacksquare$

\bibliographystyle{IEEEtran}
\bibliography{IEEEabrv,Content_Push_Ref_2017}

\end{document}